\documentclass[fleqn,usenatbib]{mnras}

\usepackage{newtxtext,newtxmath}

\usepackage[T1]{fontenc}

\DeclareRobustCommand{\VAN}[3]{#2}
\let\VANthebibliography\thebibliography
\def\thebibliography{\DeclareRobustCommand{\VAN}[3]{##3}\VANthebibliography}

\usepackage{lipsum}
\usepackage{mwe}
\DeclareUnicodeCharacter{2212}{-}

\usepackage{tabulary}

\usepackage{xcolor}

\usepackage{graphicx}	
\usepackage{amsmath}	
\usepackage{hyperref}
\hypersetup{
    colorlinks=true,
    linkcolor=blue,
    filecolor=magenta,      
    urlcolor=cyan,
}

\newcommand{\Msun}{$M_{\odot}$}

\newcommand{\kpc}{{\rm kpc}}

\newcommand{\mi}{\texttt{M12i}}
\newcommand{\midark}{\texttt{M12iDMO}}

\newcommand{\mc}{\texttt{M12c}}
\newcommand{\mcdark}{\texttt{M12cDMO}}

\newcommand{\mm}{\texttt{M12m}}
\newcommand{\mmdark}{\texttt{M12mDMO}}

\newcommand{\mf}{\texttt{M12f}}
\newcommand{\mfdark}{\texttt{M12fDMO}}

\newcommand{\mw}{\texttt{M12w}}
\newcommand{\mwdark}{\texttt{M12wDMO}}

\newcommand{\mb}{\texttt{M12b}}
\newcommand{\mbdark}{\texttt{M12bDMO}}

\newcommand{\Julietdark}{\texttt{JulietDMO}}
\newcommand{\Juliet}{\texttt{Juliet}}

\newcommand{\Romeodark}{\texttt{RomeoDMO}}
\newcommand{\Romeo}{\texttt{Romeo}}

\newcommand{\Thelmadark}{\texttt{ThelmaDMO}}
\newcommand{\Thelma}{\texttt{Thelma}}

\newcommand{\Louisedark}{\texttt{LouiseDMO}}
\newcommand{\Louise}{\texttt{Louise}}

\newcommand{\Romulusdark}{\texttt{RomulusDMO}}
\newcommand{\Romulus}{\texttt{Romulus}}

\newcommand{\Remusdark}{\texttt{RemusDMO}}
\newcommand{\Remus}{\texttt{Remus}}

\newcommand{\dd}{{\rm{d}}}

\newcommand{\spacer}{\hspace{.1cm} \vline \hspace{.1cm} }

\title[Galactic J-Factor Emission Signal Shapes]{ Analysis of shape and angular orientation parameters of velocity-dependent dark matter annihilation signals in Galactic Centers for FIRE simulations  }

\author[Daniel McKeown ]
{\parbox{17.5cm}{
Daniel McKeown$^{1}$\thanks{E-mail: dmckeown@ucla.edu}$^{}$ } \vspace{0.3cm}\\
$^{1}$University of California,Los Angeles, USA \\
}

\date{18 June 2025. March 2025;  }

\pubyear{2025}

\date{published: 18 June 2025 in Monthly Notices of the Royal Astronomical Society, staf1004, https://doi.org/10.1093/mnras/staf1004 }

\pubyear{2025}

\begin{document}
\label{firstpage}
\pagerange{\pageref{firstpage}--\pageref{lastpage}}
\maketitle

\begin{abstract}
We use FIRE-2 zoom cosmological simulations of Milky Way size galaxy halos to characterize the shape of the J-factor emission signal on the sky. We find that, at a fixed dJ/dΩ contour,the shape is well-fit by an ellipse, with semi-major axis Rmajor and semi-minor axis Rminor measured in degrees on the sky. We use least squares fitting to fit ellipses to the J-factor emission, viewed from the solar location. The ratio of minor to major axes (Rminor/Rmajor) allows us to characterize the shape of each contour, with ratio 1.0 corresponding to circular/spherical emission. These results provide new expectations for the shape of dark matter annihilation emission signals we might expect to see if dark matter is annihilating with its own anti-particle. We find that both the shape and angular orientation of the emission signal is different than results predict from dark matter only simulations. In terms of shape, we find that the ratios of semi-minor to semi-major axis is always consistently at 0.8, revealing a consistent circular emission shape, whereas with dark matter only emission signals the range of ratios from halo to halo is broader, and often closer to 0.5, showing a much more elliptical shape in general. In terms of angular orientation, we find that the major axis of the J-factor emission signal maps for FIRE halos are consistently aligned with the galactic plane within a few degrees, meaning that excess emission out of the plane would be hard to explain with a dark matter annihilation signal. However, we also find that the expected emission signal would be consistent with Fermi-LAT measurements showing a galactic center excess in full hydrodynamical simulations with stars and gas included, while dark-matter-only simulations do not produce the expected signal.

\end{abstract}

\begin{keywords}
galaxies: disc -- galaxies: formation -- cosmology: dark matter -- cosmology: theory
\end{keywords}



\section{Introduction}

An extended source of gamma-ray emission has been detected by Fermi-LAT telescope \cite{Hooper11,murgia2020,Giacchino13,Choquette16,Petac18,Boddy18,johnson2019search,Arguelles19,Board21}.
The shape of this emission on the sky is an important ingredient for determining the source(s).  These could include milli-second pulsars(MSPs), hot gas and plasma in the disk, supernovae events, and potentially dark matter annihilation (see  \cite{Abazajian20} for a discussion).   

Although an excess has been detected, significant challenges exist in resolving the data (see \cite{ackermann2012fermi}). Excess gamma ray flux has been seen in the direction of M31 as well, \cite{zimmer2022andromeda}. This shows that the effect is not isolated to our own MW Galaxy. In the M31 case, there are uncertainties in the astrophysical gamma-ray foreground model \cite{zimmer2022andromeda} that make it challenging to determine specifically the source of the gamma-rays. If we consider MSPs as a potential source of the emission signal, there is a quantitative way to decide how large of a population of millisecond pulsars should exist in the overall population of stars (see for example \cite{winter2016estimating} who did this for the case of Dwarf Spheroidal Galaxies ). 

However, there is currently only a small number of  MSPs that have been determined to exist in the MW disk (approx. 100 MSPs)\cite{winter2016estimating}. Thus, such analysis involves assigning a population of hypothetical MSPs using scaling functions based on the stellar masses of the population for the region in question \cite{winter2016estimating,Dinsmore22} Interpreting the magnitude of the signal at the galactic center as due to MSPs is left to being approximate and not definitive. It is difficult then to definitely rule out in favor for or against the gamma rays being from another source ( see  \cite{kimura2021soft} and \cite{tominaga2007connection}) However, if additional criteria are examined, such as the shape of the emission signal, there could be a more definitive answer \cite{winter2016estimating},\cite{zimmer2022andromeda}, \cite{Abazajian20},\cite{kimura2021soft}, \cite{tominaga2007connection}. 

As an example of the power of emission-shape templates, 
 \cite{Abazajian20} showed that allowing for a  "boxy" bulge template for stellar emission allowed them to rule out a significant range of thermal WIMP dark matter models.
Interestingly, however, \cite{di2021characteristics} found that the Galactic Center Excess has a shape on the sky well fit by an ellipsoid with a fairly round axis ratio $0.8 - 1.2$, where the axes are aligned along the galactic plane.  This is more spherical than the ``boxy bulge" shape of emission reported in Ref. \cite{Macias18} (which has a short-to-long ratio of $\sim 0.55$) and used by \cite{Abazajian20} to rule out some thermal dark matter models.  Instead, \cite{di2021characteristics} conclude that the more spherical GCE shape is more consistent with what is expected from a dark matter annihilation signal.  They suggest the disagreement is owing to the different background components used in the two analyses.
Independent of who is ultimately correct in characterizing the shape of the galactic center excess, the expected shape of dark matter emission  is key to the final interpretation.

Similarly, \cite{zimmer2022andromeda} have utilized the shape in the case of M31 using the PAndAS survey and removal of foreground stars to map the stars in the Galactic region of interest and then assign a probabilistic population of millisecond pulsars in the bulge. They found that this population would explain the gamma ray excess without the need for any dark matter annihilation. This was done with modeling the stellar population and its distribution. While \cite{zimmer2022andromeda} took into account the shape of the stellar populations in  their analysis, they assumed that the dark matter halo was spherical. It is possible that with a careful allowance for realistic, non-spherical dark matter structure around M31 that the dark matter interpretation could be warranted.  Knowing the potential shape of such an emission signal based on its location is thus useful in itself for future studies that may rule out or limit theories for dark matter annihilation as a possible gamma ray source. 


Typically, authors assume analytic dark matter profiles when interpreting indirect detection signals.  Popular models include spherical NFW, spherical cored profiles, and/or 3D ellipsoidal versions of those profiles, with specific choices of axis ratios   (see \cite{Abazajian20}). Numerical simulations provide a more precise prediction for the expected shapes of dark matter halos,  including  \cite{bernal2016spherical} who use Illustris simulations, with somewhat lower resolution than our own, and conclude that galaxy formation renders halos more spherical, on average, that DMO simulations. 

In what follows, we examine the shape of the expected dark matter annihilation emission signal using J-factor sky maps of the simulations described in \cite{2022MNRAS.513...55M}. Rather than estimate the 3d shape of our dark matter distributions with approximate ellipsoidal fits, we instead fit ellipses on the sky from mock observer planes. This allows a more direct prediction for what will be observed without the need to project the 3d fits.  Our goal is to provide accurate dark-matter annihilation shape estimates, in order to enable studies to distinguish excess emission from the many other possible sources.

Past simulation work shows us that dark matter emission is somewhat circular on the sky. For example, \cite{bernal2016spherical} used hydrodynamic simulations, with somewhat lower resolution than our own, to show through their analysis that s-wave annihilation signals from dark matter sources should be fairly symmetrical with axis ratios typically greater than $0.8$. It is important to note that they defined these ratios in terms of a J-factor-weighted-inertia-tensor, averaged over the whole sky. Our work, which allows us to track emission shapes quite close to the Galactic Center (within $\sim 3^\circ$), allows us to measure the shape as a function of angle on the sky and we extend the shape analysis to p-wave and d-wave models as well. The reason we do this, is because it is closer to how an observational signal is measured.

Furthermore, \cite{grand2022dark} have found that in high resolution galactic simulations, the concentration of dark matter particles for full hydrodynamics is greater when including the effects of gas and star particles, and that this results in an emission signal that could be consistent with the Galactic Center Excess observed in FermiLAT data. Our results in this paper go beyond this to show what the signal would look like in terms of shape, while also revealing that this effect was due to baryonic contraction, and that this resulted in a more centrally concentrated dark matter density profiles at the stellar mass scales. We originally found this in a previous paper (see Figure 1) from \cite{2022MNRAS.513...55M}. However in the previous paper the exact shape of the signal found was not calculated as it is in this current one. Similar to  \cite{grand2022dark}, we have found that the for s-wave models, the emission signal we observe for Full Baryonic runs of halos is more morphologically consistent with the Galactic Center Excess than DMO runs. In addition to what  \cite{grand2022dark}have found, we also have observed this for p and d wave models as well. This is due to the suppression of substructure due to baryonic contraction in these velocity dependent models.

If baryonic contraction is occurring for dark matter particles at the galactic center of the Milky Way, this could be responsible for the emission signal observed by FermiLAT. However, more careful analysis is needed, as noted by \cite{grand2022dark}, in order to differentiate between this and other possible sources of the excess in FermiLAT data. One of the best ways to differentiate whether the signal is due to a possible source of dark matter annihilation or not is a detailed look at the shape of the emission signal itself. This is the primary goal of this paper, where we compare the shape of the expected emission signal from dark matter only and fully hydrodynamical simulations, and characterize the shape in a fully consistent way so that it can be compared with the shape of FermiLAT data.

\cite{piccirillo2022velocity} have examined the velocity dependent J-factors for p and d wave, as well as Sommerfeld models using the Auriga simulations. While their work finds intriguing connections between the emission signal and the effects of velocities for dark matter particles, they do not have the same ability to produce detailed contour maps to the scale that we have been able to achieve. Due to the high resolution of Fire Simulations, we have resolved to within 2.75 deg compared to 10 (APOSTLE) deg and 7 (Auriga) deg. \cite{2022MNRAS.513...55M} This very high degree of resolution we have achieved has thus motivated both a detailed analysis of the overall magnitude of the J-factor emission for s,p and d wave models as well as looking at the shape and degree of elongation of the signals for both dark matter only and full hydrodynamics simulations. Thus, we have a way to test more parameters of the fit of the galactic center excess than previously conducted analysis.


\begin{table*}
	\centering
	\caption{(1) Simulation name. The suffix ``DMO" stands for ``Dark Matter Only" and refers to the same simulation run with no hydrodynamics or galaxy formation physics. (2) Factor $f$ by which dark matter particle masses have been multiplied ($m_{\rm dm} \rightarrow fm_{\rm dm}$) in order to normalize the dark matter density at $\rho(r = R_\odot) = 10^7$ \Msun  $\kpc^{-3} =  0.38 $GeV cm$^{-3}$ for a mock solar location $R_\odot = 8.3$ kpc. (3) Stellar mass $M_\star$ of the central galaxy. (4) Virial mass (of raw simulation, not including the $f$ factor) defined by \citet{ByranNorman1998}. On-sky ellipse fits to s-wave j-factor contours of fixed value $10^{24} (Gev)^{2}cm^{-5}Sr^{-1}$  ($*$ implies $10^{23.7} (Gev)^{2}cm^{-5}Sr^{-1} $). Columns list the following:  (5) The semi-major axis of the elliptical fit  (6) The semi-minor axis of the elliptical fit. (7) The ratio of semi-minor to semi-major axis. (8-10) On-sky ellipse fits to p-wave contours of fixed value $ 10^{17.5} (Gev)^{2}cm^{-5}Sr^{-1}$ .  Note $*$ implies $ 10^{16.9}(Gev)^{2}cm^{-5}Sr^{-1}$ in the same units and  $**$  implies $ 10^{17.1} (Gev)^{2}cm^{-5}Sr^{-1} $) columns are same as for s-wave runs. (11-13) On-sky ellipse fits to d-wave contour values for m12 runs at values of $10^{11.5} (Gev)^{2}cm^{-5}Sr^{-1}$.  Note $*$ implies $ 10^{10.8} (Gev)^{2}cm^{-5}Sr^{-1} $  ,  $**$ implies $ 10^{11.8} (Gev)^{2}cm^{-5}Sr^{-1}$ 
	   }
        \label{tab:one}
	\begin{tabular}{lccc @{\spacer} ccc @{\spacer} ccc @{\spacer} ccc} 
		\hline
		Simulation & f & M$_\star$ & M$_{\rm vir}$ &  s-wave &  &  & p-wave &  &  & d-wave &  &  \\ 
		           &   & $10^{10}$ M$_\odot$ &  $10^{12}$ M$_\odot$ & 
		             &  $\hspace{-2.2 cm}$ ($10^{24}$ & \hspace{-2 cm}   $(Gev)^{2}cm^{-5}$) & &  $\hspace{-2.2 cm}$ ($10^{16}$ & \hspace{-2 cm}   $(Gev)^{2}cm^{-5}$) &  &  $\hspace{-2.3 cm}$ ($10^{10}$ & \hspace{-2 cm}   $(Gev)^{2}cm^{-5}$) \\  
		
                      &   &  
		                &  & major &  minor &  ratio & major & minor  & ratio  &  major & minor & ratio  \\

            \hline
		\mi         & 1.28 & 6.4  & 0.90   & *$11.2^{ \circ }$  & * $8.95^{ \circ }$  & 0.8  &  *$21.1^{ \circ }$ &   *$16.6^{ \circ }$ & 0.79   & **$22.6^{ \circ }$  &  ** $17.7^{ \circ }$ & 0.79  \\
		
        \midark     & 1.59 & -   & 1.3  &* $4.41^{ \circ }$   & *$2.92^{ \circ }$  & 0.66  &  $*11.1^{ \circ }$  & $*6.04^{ \circ }$  & 0.54  &  *$5.77^{ \circ }$  & *$2.10^{ \circ }$ & 0.36   \\ \\
		
		\mc         & 1.26  & 6.0  & 1.1  & $6.18^{ \circ }$  &  $4.21^{ \circ }$   & 0.71      &  $ *33.3^{ \circ }$ & *$22.6^{ \circ }$ & 0.67  & *$44.0^{ \circ }$  &   *$31.4^{ \circ }$  & 0.71    \\
		\mcdark     & 1.83  & -  & 1.3  &  $6.52^{ \circ }$ &  $3.21^{ \circ }$ & 0.49   &  *$15.6^{ \circ }$ &  *$8.03^{ \circ }$ & 0.52   & *$10.9^{ \circ }$ &  *$6.64^{ \circ }$ & 0.61   \\ \\
		
		\mm         &  0.885 &  11 & 1.2  & $3.91^{ \circ }$ &  $1.49^{ \circ }$ & 0.38   & $24.3^{ \circ }$ &  $15.6^{ \circ }$ & 0.64  &  $38.1^{ \circ }$ & $24.8^{ \circ }$  & 0.65 \\
		\mmdark    &  1.42  &  - & 1.4 &  $6.23^{ \circ }$   & $4.13^{ \circ }$ & 0.66  &  $6.49^{ \circ }$ &  $4.42^{ \circ }$ & 0.68  &  *$15.5^{ \circ }$ & *$12.3^{ \circ }$ & 0.80  \\ \\
		
		\mf         &  1.01  &  8.6 & 1.3 & $5.42^{ \circ }$ & $4.01^{ \circ }$ & 0.74 &   $21.9^{ \circ }$  & $17.3^{ \circ }$  & 0.79   & $27.1^{ \circ }$ & $21.2^{ \circ }$ & 0.78    \\
		\mfdark    &  1.82  &  - &  1.6 &  $5.53^{ \circ }$   & $3.17^{ \circ }$ & 0.57  &  $3.82^{ \circ }$  &  $2.30^{ \circ }$  & 0.60  & $8.31^{ \circ }$  & $5.87^{ \circ }$ & 0.71  \\ \\
		
		\mw        &  1.28  &  5.8  &  0.83  &  $5.08^{ \circ }$ &  $4.21^{ \circ }$ & 0.83  &  **$30.0^{ \circ }$ & ** $23.0^{ \circ }$ & 0.77 & $28.0^{ \circ }$ & $20.6^{ \circ }$ & 0.74  \\
		\mwdark    &  1.68  &  - & 1.1 &  $2.71^{ \circ }$ & $1.96^{ \circ }$  &  0.72  &  **$8.05^{ \circ }$  &  ** $5.68^{ \circ }$  & 0.70   & *$8.93^{ \circ }$ &  *$6.00^{ \circ }$ & 0.67\\ \\
		
		\mb          &  0.990 &  8.1  &  1.1  & $6.61^{ \circ }$ & $5.24^{ \circ }$ & 0.79  & $23.2^{ \circ }$ &  $19.5^{ \circ }$ & 0.84  & $33.1^{ \circ } $ & $27.9^{ \circ }$  & 0.84 \\
		\mbdark      &  1.25  &  -  & 1.4 & $5.73^{ \circ }$ & $5.46^{ \circ }$  & 0.95  & $7.94^{ \circ }$ & $7.42^{ \circ }$ & 0.93  & $5.19^{ \circ }$ & $4.93^{ \circ }$ & 0.95  \\  \\

		\Romeo      & 0.99  & 7.4  & 1.0 & $10.5^{ \circ }$ & $7.70^{ \circ }$ & $0.74$  &  $23.6^{ \circ }$ & $17.3^{ \circ }$ & 0.73  & $31.7^{ \circ }$ & $23.8^{ \circ }$ & 0.75   \\
		\Romeodark  & 1.26  & - & 1.2 & $7.0^{ \circ }$  &  $5.6^{ \circ }$ &  0.80  &  $8.83^{ \circ }$ & $7.39^{ \circ }$ & 0.84  &  $6.50^{ \circ }$ & $5.20^{ \circ }$ & 0.80 \\ \\
	
			\Juliet     & 1.31  & 4.2  & 0.85  &  $9.50^{ \circ }$ & $8.13^{ \circ }$ & 0.86   & $18.5^{ \circ }$ & $15.6^{ \circ }$ & 0.84  & $23.0^{ \circ }$ & $19.6^{ \circ }$   & 0.85 \\
		\Julietdark & 1.59  & -    &  1.0   & $8.61^{ \circ }$ & $6.04^{ \circ }$ & 0.70 &   $10.8^{ \circ }$ & $7.53^{ \circ }$ & 0.70   & $5.90^{ \circ }$ & $4.67^{ \circ }$ & 0.80   \\ \\

		\Thelma      & 1.17 & 7.9 & 1.1 &  *$2.97^{ \circ }$ & *  $2.26^{ \circ }$  &  0.76  &   $18.14^{ \circ }$  & $13.9^{ \circ }$ &  0.77  & **$20.5^{ \circ }$  & **$16.6^{ \circ }$ & 0.81 \\
		\Thelmadark  & 1.70  & -    & 1.3  & $6.91^{ \circ }$ &  $2.75^{ \circ }$ &  0.40  & $4.53^{ \circ }$  & $1.77^{ \circ }$   &  0.39   & *$16.7^{ \circ }$ & *$6.98^{ \circ }$ & 0.42  \\  \\

		\Louise     &  1.42  & 2.9    & 0.85  & $11.4^{ \circ }$  &  $8.85^{ \circ }$ & 0.77  &  $16.9^{ \circ }$ & $12.9^{ \circ }$ & 0.76   & $19.9^{ \circ }$&  $15.6^{ \circ }$ & 0.79  \\
		\Louisedark  & 1.41  & -    & 1.0  & $5.59^{ \circ }$ & $4.83^{ \circ }$ & 0.86   &  $7.60^{ \circ }$ &  $6.74^{ \circ }$ & 0.89   & $4.0^{ \circ }$  & $2.8^{ \circ }$ & 0.69  \\  \\

		\Romulus     & 1.00  & 10   & 1.53 & $13.3^{ \circ }$ & $10.2^{ \circ }$ & 0.77  &  $24.0^{ \circ }$ & $18.0^{ \circ }$  & 0.75  & $35.5^{ \circ }$ & $26.6^{ \circ }$ & 0.75 \\
		\Romulusdark  &  1.01   & -    & 1.9 &  $5.44^{ \circ }$ &  $3.07^{ \circ }$  & 0.56   &  $8.46^{ \circ }$ & $5.03^{ \circ }$ & 0.59   &  $5.81^{ \circ }$ &  $3.13^{ \circ }$ & 0.54  \\ \\

		\Remus    & 1.10  & 5.1  & 0.97 & $12.3^{ \circ }$ & $9.34^{ \circ }$ & 0.76  &   $20.5^{ \circ }$ &  $15.7^{ \circ }$ & 0.77 & $27.1^{ \circ }$ & $21.2^{ \circ }$ & 0.78 \\
		\Remusdark  & 1.18  & -  & 1.3 & $7.65^{ \circ }$ & $5.00^{ \circ }$ & 0.65   &  $10.7^{ \circ }$ & $7.35^{ \circ }$ & 0.69    & $8.31^{ \circ }$ & $5.87^{ \circ }$ & 0.71   \\ \\

		\hline
	\end{tabular}
\end{table*}

\section{Overview of Simulations}

Our analysis relies on cosmological zoom-in simulations performed as part of the Feedback In Realistic Environments (FIRE) project\footnote{\url{https://fire.northwestern.edu/}} with FIRE-2 feedback implementation \citep{Hopkins17} with the gravity plus hydrodynamics code {\small GIZMO} \citep{Hopkins15}.  FIRE-2 includes radiative heating and cooling for gas with temperatures ranging from 10\,{\rm K} to $10^{10}\,{\rm K}$, an ionising background \citep{Faucher2009}, stellar feedback from OB stars, AGB mass-loss, type Ia and type II supernovae, photoelectric heating, and radiation pressure. Star formation occurs in gas that is locally self-gravitating, sufficiently dense ($ > 1000$ cm$^{-3}$), Jeans unstable, and molecular (following \citealt{Krumholz_2011}). Locally, the star formation efficiency is set to $100\%$ per free-fall time, though the global efficiency of star formation within a giant-molecular cloud (or across larger scales) is self-regulated by feedback to $\sim$1-10\% per free-fall time \citep{Orr_2018}.

In this work, we analyse 12 Milky-Way-mass galaxies (Table \ref{tab:one}). These zoom simulations are initialised following the approach outlined in \citet{Onorbe14} using the MUSIC code \citep{HA11}. Six of these galaxies were run as part of the Latte suite ~\citep{Wetzel16,Garrison-Kimmel17,Garrison-Kimmel19,samuel2020profile,Hopkins17} and have names following the convention \texttt{m12*}. The other six, with names associated with famous duos, are set in paired configurations to mimic the Milky Way and M31 \citep{Garrison-Kimmel19,Garrison-Kimmel19_2}. Analysis has shown these are good candidates for comparison with the Milky Way \citep{sanderson2020synthetic}. Gas particles for the \texttt{M12*} runs have initial masses of $m_{\rm g,i} = 7070\,$ \Msun.  The ELVIS on FIRE simulations have roughly two times better mass resolution ($m_{\rm g,i} \simeq 3500 - 4000$ \Msun). Gas softening lengths are fully adaptive down to $\simeq$0.5$-$1 pc. The dark matter particle masses are $m_{\rm dm} = 3.5 \times 10^4$ \Msun for the Latte simulations and $m_{\rm dm} \simeq 2 \times 10^4$ \Msun for the ELVIS runs.  Star particle softening lengths are $\simeq$4 pc physical and a dark matter force softening is $\simeq$40 pc physical. These particle masses translate to a higher resolution than similar simulations. For example, to within 10 degrees for (APOSTLE) and 7 (Auriga) deg, compared to our 2.75 deg resolution.

Substructures within the halos are apparent. Although we choose to focus on the galactic center, and did not focus on characterizing the influence of this substructure, the high resolution allows us to resolve substructures that are within approximately ~ 5 percent of the total virial mass of the dominant halo. Stellar feedback reduces substructures, so that dark matter only runs have more substructure than full baryonic simulations.

In the skymaps which follow, substructure is difficult to discern from the background noise. We have chosen to focus only on the dominant halo of the galactic center in this study which seeks to characterize the galactic center excess. Thus, to see the effect of substructure on the total signal, more analysis would be needed in the future. However, the contribution to the J-factor for substructure would presumably be much smaller for substructures, given that the primary galactic center contains far more dark matter in which to annihilate into gamma ray photons than any smaller substructures. Also, the angular focus is within 45 degrees in any direction from the galactic center, so many substructures that are contained within the simulation that are farther out from the center would not be seen or taken as part of the J-factor calculation. 

Lastly, each FIRE simulation has an analogous dark matter only (DMO) version. The individual dark matter particle masses in the DMO simulations are larger by a factor of $(1 − f_{\rm b})^{−1}$ in order to keep the total gravitating mass of the Universe the same, where  $f_{\rm b} = \Omega_{ \rm b} /\Omega_{\rm m}$ is the cosmic baryon fraction.  The initial conditions are otherwise identical. DMO versions of each halo are referred to with the same name as the FIRE version with the added suffix ``DMO."

As can be seen in Table \ref{tab:one}, the stellar masses of the main galaxy in each FIRE run (second column) are broadly in line with the Milky Way: $M_\star \approx (3 - 11) \times 10^{10}$ \Msun.  
The virial masses \citep{ByranNorman1998} of all the halos in these simulation span a range generally in line with expectations for the Milky Way: $M_{\rm vir} \approx (0.9 - 1.8) \times 10^{12} $ \Msun.   In every case, the DMO version of each pair ends up with a higher virial mass.  This is consistent with the expectation that halos will have lost their share of cosmic mass by not retaining all baryons in association with feedback. As we discuss in the next section, in our primary analysis we re-normalize all halos (both FIRE and DMO runs) so that they have the same ``local" dark matter density at the Solar location (by the factor $f$ listed in the table).

\section{Astrophysical J-Factors}

\subsection{Definitions}

If dark matter particle of mass $m_\chi$ is its own antiparticle with an annihilation cross section $\sigma_A$, the resulting differential particle flux produced by annihilation in a dark matter halo 
can be written as the integral along a line of sight $\ell$ from the observer (located at the solar location in our case) in a direction $\vec{\theta}$ in the plane of the sky over pairs of dark matter particles with velocities $\vec{v}_1$ and $\vec{v}_2$:
\begin{equation}
  \frac{d^2\Phi}{dE d\Omega} =
  \frac{1}{4\pi} \frac{d N}{d E} \int \dd \ell ~\dd^3v_1 \, \dd^3 v_2 
   \frac{f(\vec{r}, \vec{v}_1)}{m_\chi}
  \frac{f(\vec{r}, \vec{v}_2)}{m_\chi}
  \, \frac{(\sigma_A v_{\rm rel}) }{2} \ .
\end{equation}
Here, $\vec{r} = \vec{r}(\ell,\vec{\theta})$ is the 3D position, which depends on the distance along the line of sight $\ell$ and sky location $\vec{\theta}$.  The dark matter velocity distribution $f(\vec{r}, \vec{v})$  is normalized such that $\int d^3v f(\vec{r}, \vec{v}) = \rho (\vec{r})$, where $\rho$ is the dark matter density at that location.  The symbol $v_{\rm rel}=|\vec{v}_1 - \vec{v}_2|$ represents the relative velocity between pairs of dark matter particles.  The quantity $m_\chi$ is the dark matter particle mass and $d N / d E$ is the particle energy spectrum ultimately produced by a single annihilation.  

Following \citet{Boddy18}, we parameterize the velocity-dependence of the dark matter annihilation cross section as 
\begin{equation}
    \sigma_A v_{\rm rel} = [\sigma v]_0 \, Q(v_{\rm rel}),
    \label{eq:sigv}
\end{equation}
where $[\sigma v]_0$ is the overall amplitude and the function $Q(v)$ parameterizes the velocity dependence. For $s$-wave, $p$-wave, and $d$-wave annihilation, $Q(v) = 1$, $(v/c)^2$, and $(v/c)^4$, respectively.  We can then  rewrite the differential particle flux as
\begin{equation}
  \frac{d^2 \Phi}{d E \dd \Omega} =
  \frac{(\sigma_A v)_0}{8\pi m_X^2} \frac{d N}{d E_\gamma} \left[ \frac{d J_Q}{d  \Omega} \right]  \, .
\label{eq:Jfactor}
\end{equation}
Here, the term in brackets absorbs all of the astrophysics inputs and defines the astrophysical "J-factor" 
\begin{equation}
  \frac{dJ_Q}{d \Omega} (\vec{\theta}) = 
  \int \dd \ell \int \dd^3 v_1 {f(\vec{r}, \vec{v}_1)}
  \int \dd^3 v_2 {f(\vec{r}, \vec{v}_2)}\, Q( v_{\rm rel}) \, .
  \label{eq:Jfactor_def}
\end{equation}
In principle, the $\ell$ integral above sums pairs along the line-of-sight from the observer ($\ell = 0$) to infinity. In practice, we are focusing on J-factors arising from an individual ``Milky Way" halo, and truncate our integrals at the halo's edge (see below).

It is often useful to quote the cumulative J-factor within a circular patch of sky of angular radius $\psi$ centered on the Galactic Center.  In this case, the patch defined by $\psi$ subtends a solid angle $\Omega_\psi = 4 \pi \sin^2({\psi}/2)$ and we have:
\begin{equation}
  J_Q (< \psi)   =  \int_0^{\Omega_\psi} \frac{d J_Q}{d  \Omega} (\vec{\theta}) \,  \dd \Omega \, .
  \label{eq:Jf}
\end{equation}

\begin{figure*}
	\includegraphics[width=\columnwidth,trim = 130 0 50 90]{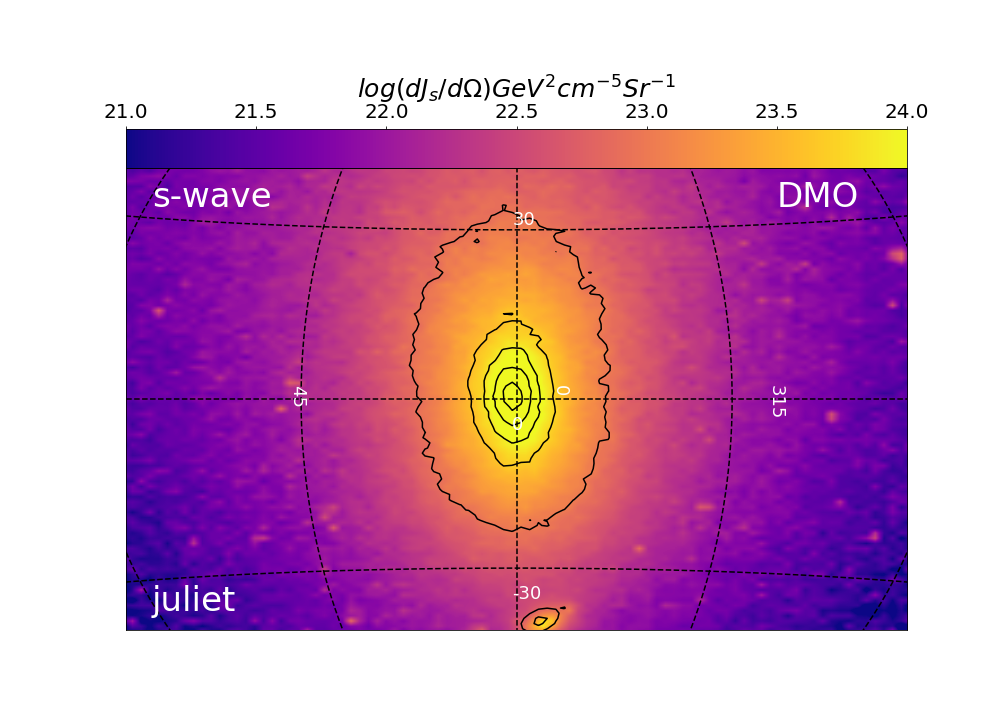}
		\includegraphics[width=\columnwidth, trim = 50 0 130 90]{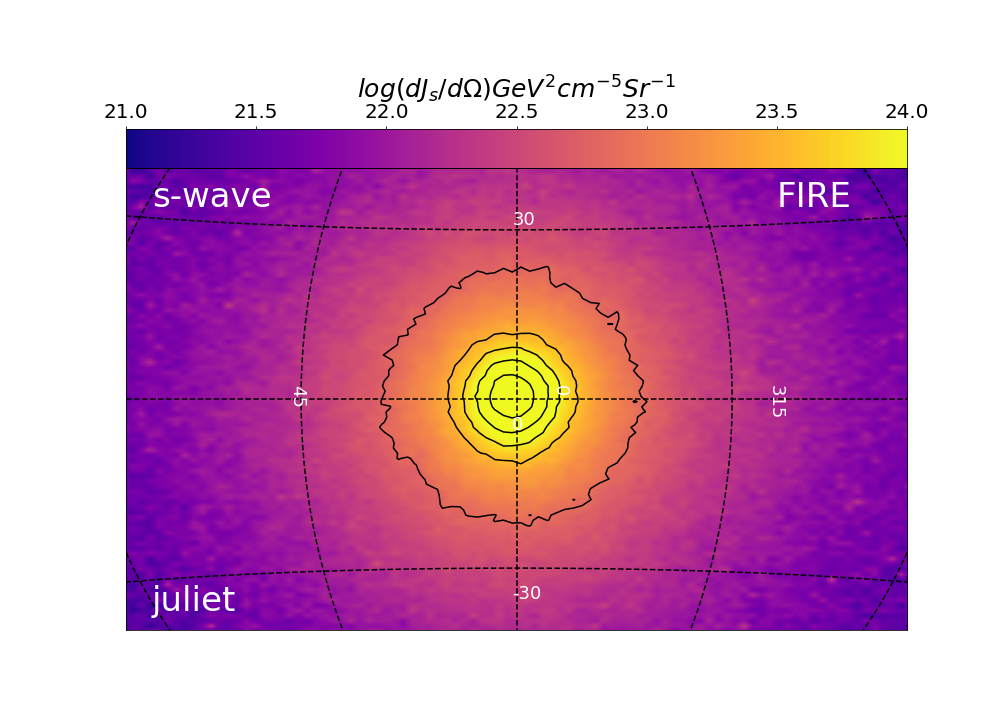} \\
	\includegraphics[width=\columnwidth,trim = 130 0 50 90]{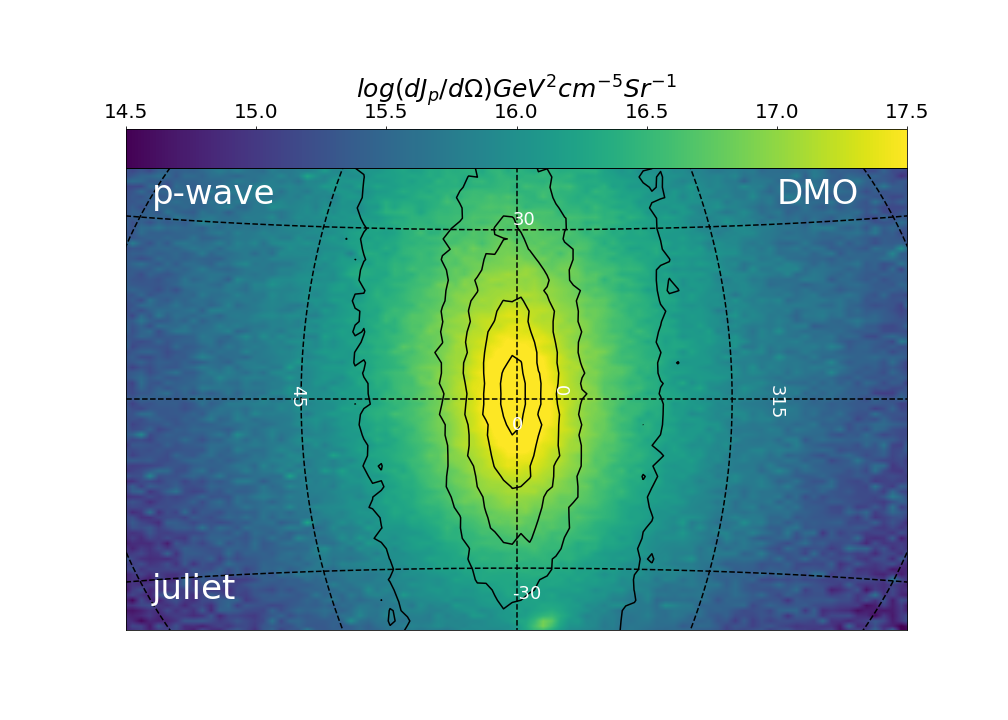}
		\includegraphics[width=\columnwidth, trim = 50 0 130 90]{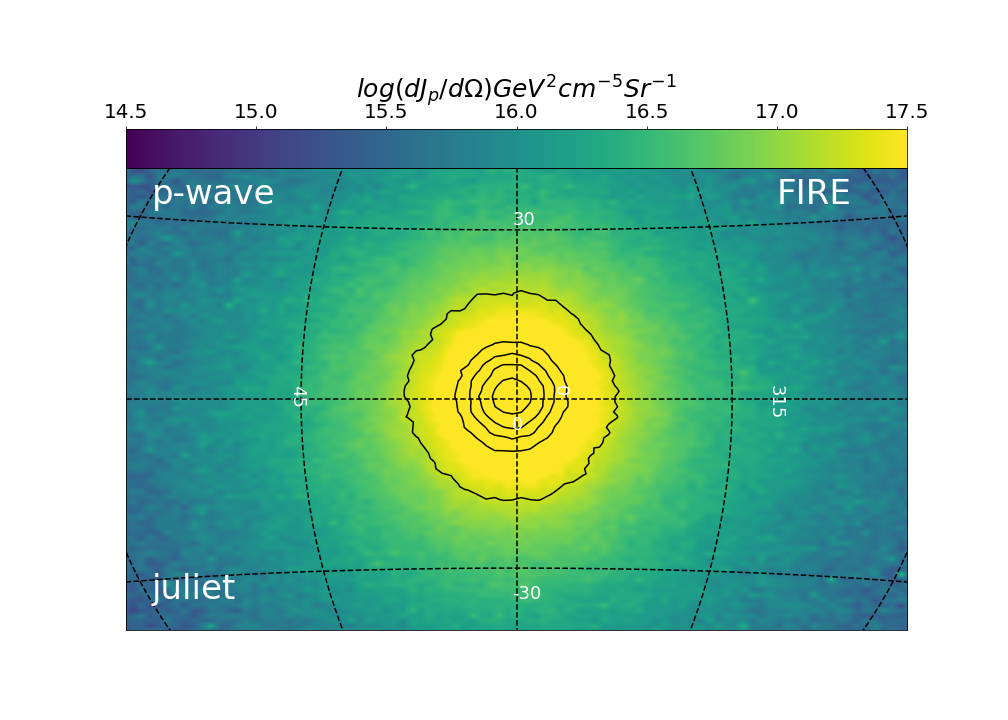} \\
	\includegraphics[width=\columnwidth,trim = 130 0 50 90]{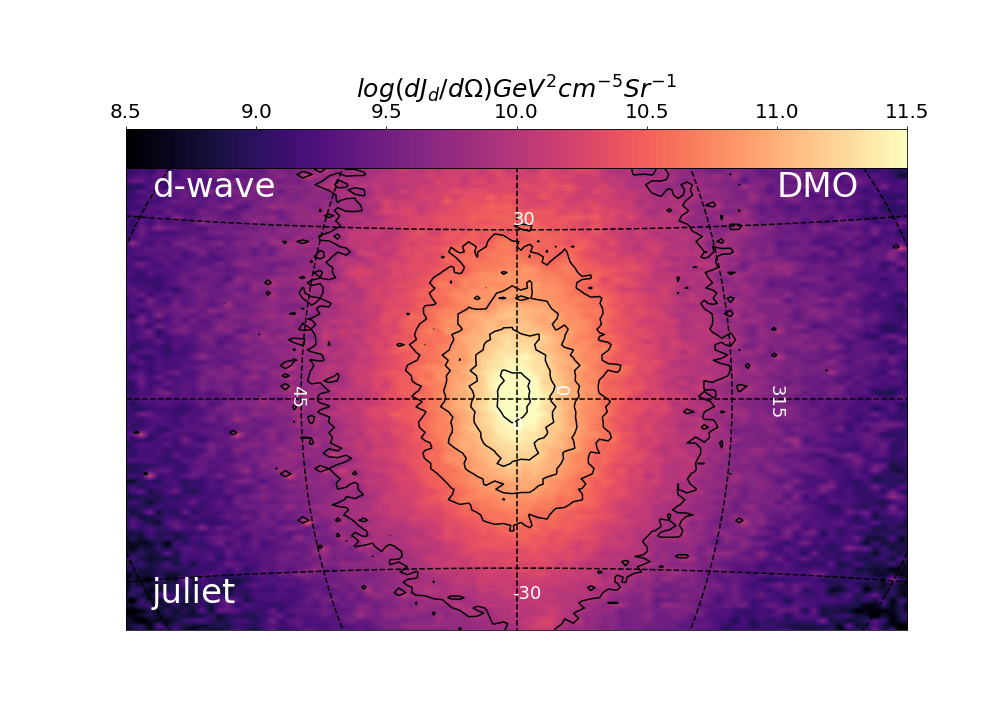}
		\includegraphics[width=\columnwidth, trim = 50 0 130 90]{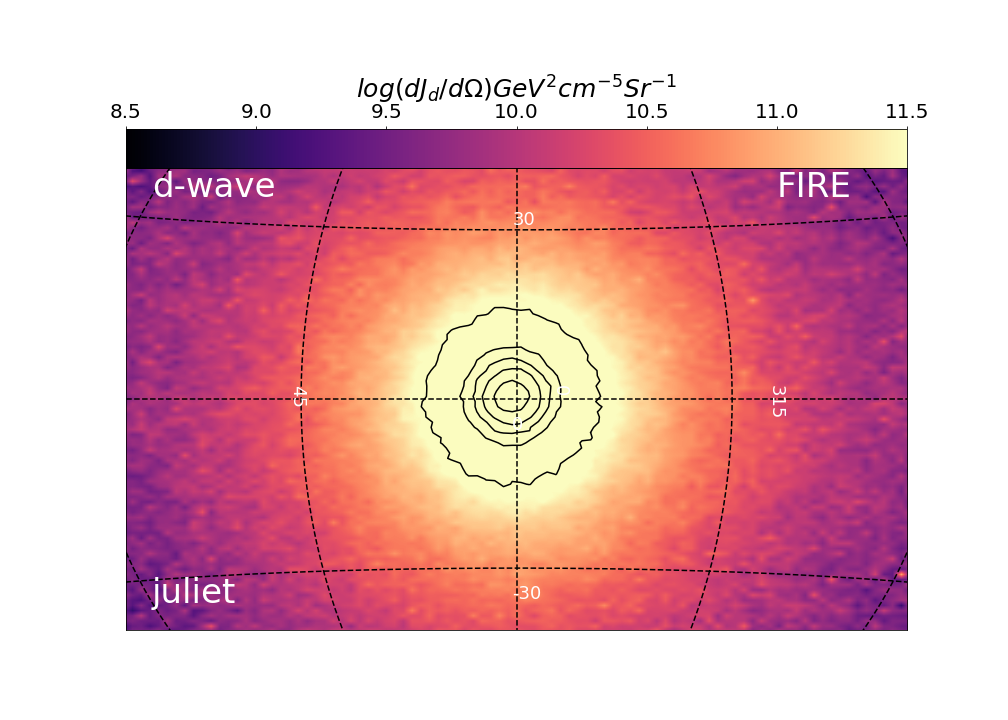}
    \caption{All sky maps of s-wave(top), p-wave(middle) and d-wave(bottom) J-factor for \texttt{Juliet}.  The  DMO version is shown on the left and full physics FIRE on the right. The horizontal and vertical lines are spaced by 30 and 45 degrees respectively, to guide the eye. To emphasize the shape of the emission, black solid contours are drawn at a fixed fraction of the maximum pixel in each halo: 0.5, 0.2, 0.1, 0.05, and 0.01. Note that FIRE contours are more symmetric and circular than the DMO contours. Note the substructure in DMO maps. We see some mild flattening along the disk plane in the FIRE run.}
    \label{fig:map_juliet}
    
\end{figure*}

\begin{figure*}
	\includegraphics[width=\columnwidth,trim = 130 0 50 90]{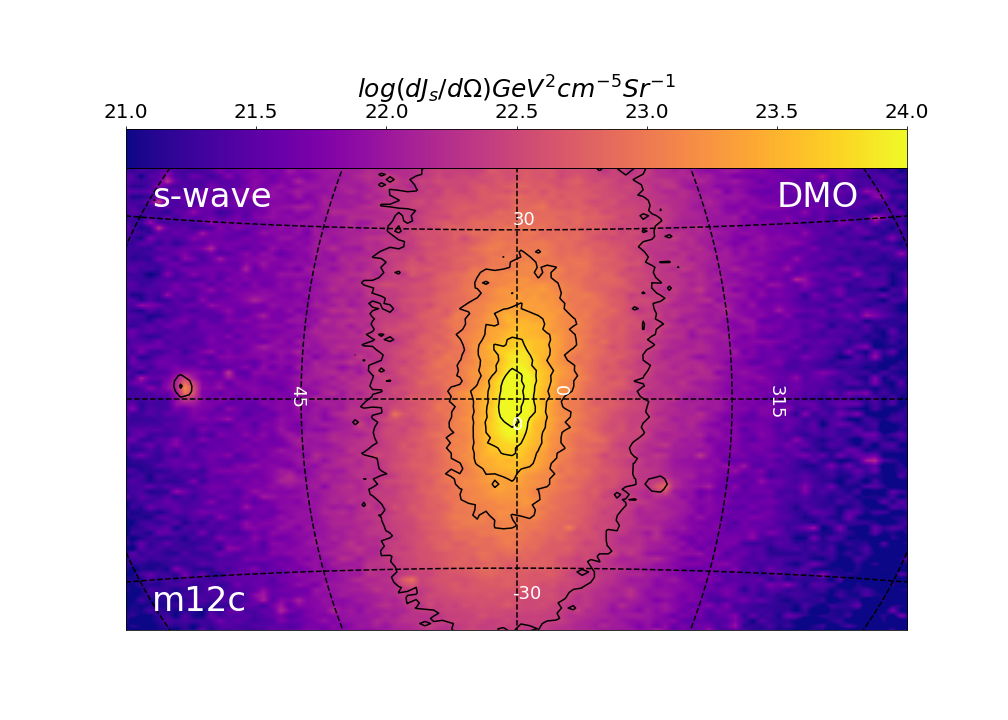}
		\includegraphics[width=\columnwidth, trim = 50 0 130 90]{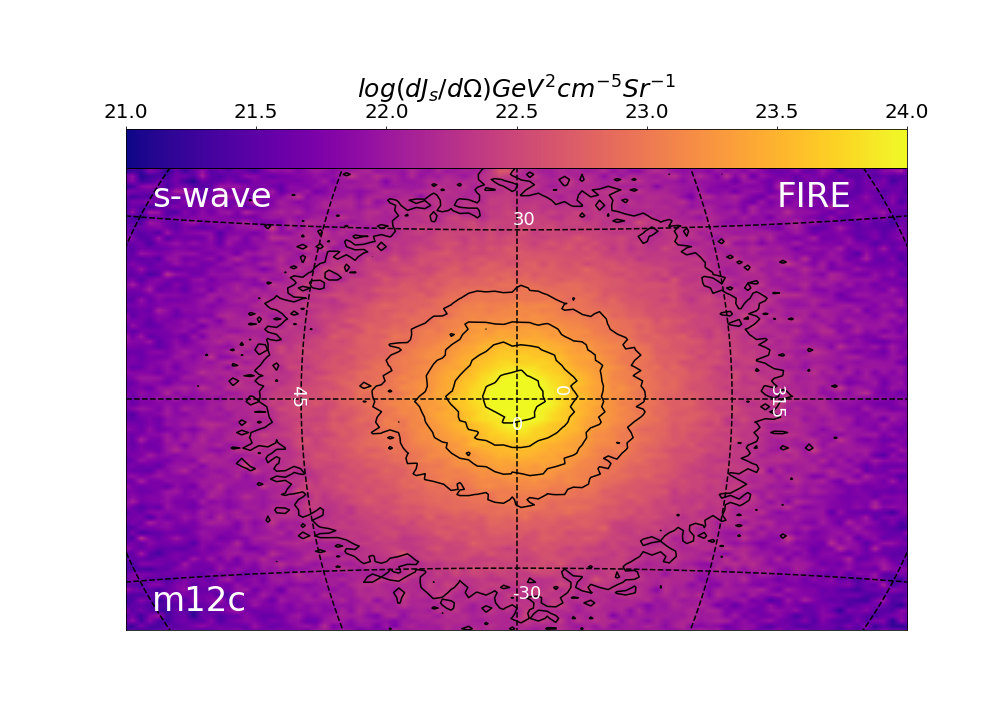} \\
	\includegraphics[width=\columnwidth,trim = 130 0 50 90]{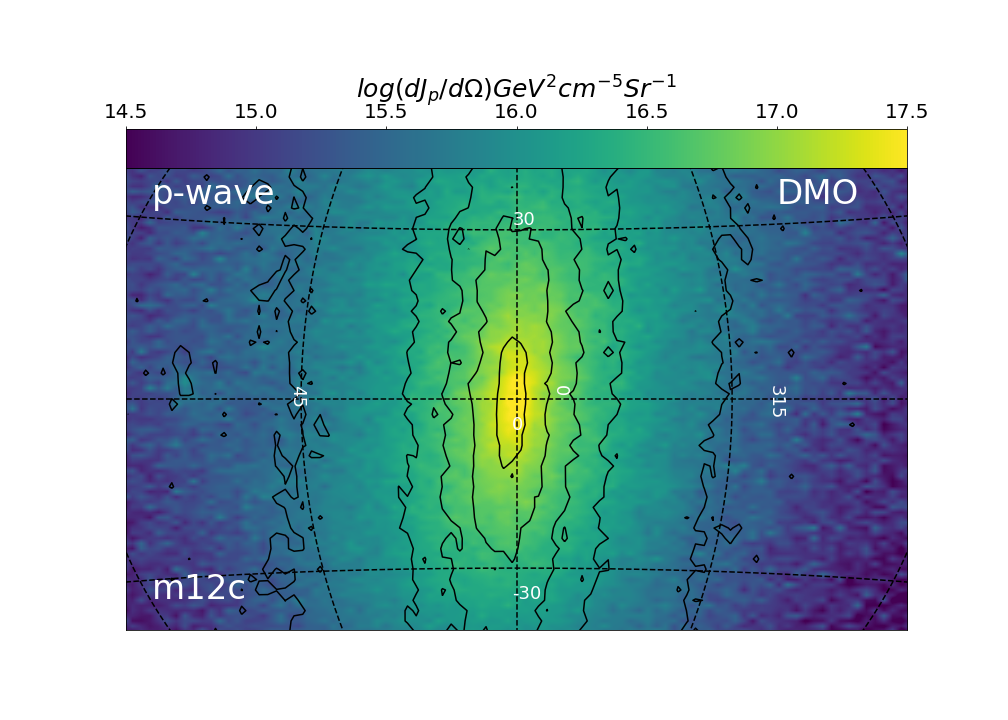}
		\includegraphics[width=\columnwidth, trim = 50 0 130 90]{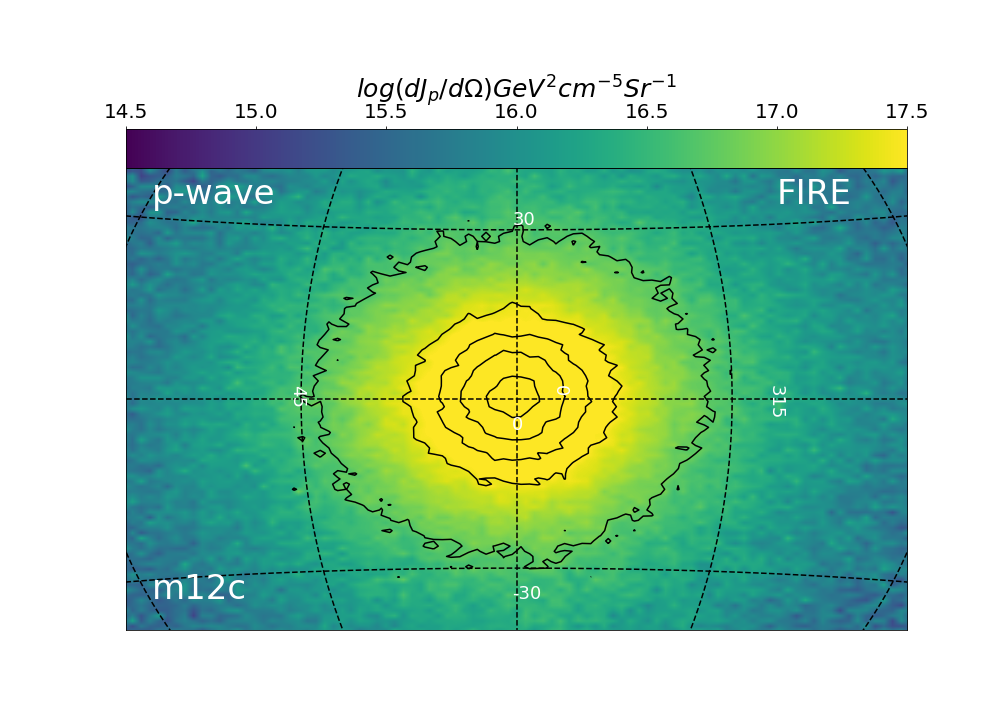} \\
	\includegraphics[width=\columnwidth,trim = 130 0 50 90]{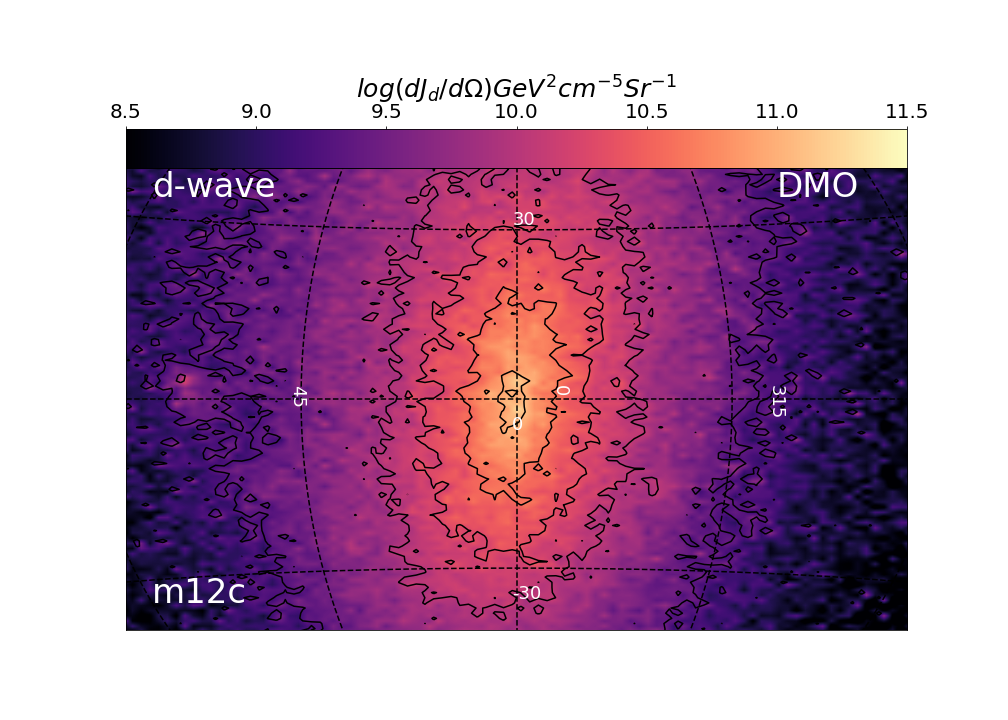}
		\includegraphics[width=\columnwidth, trim = 50 0 130 90]{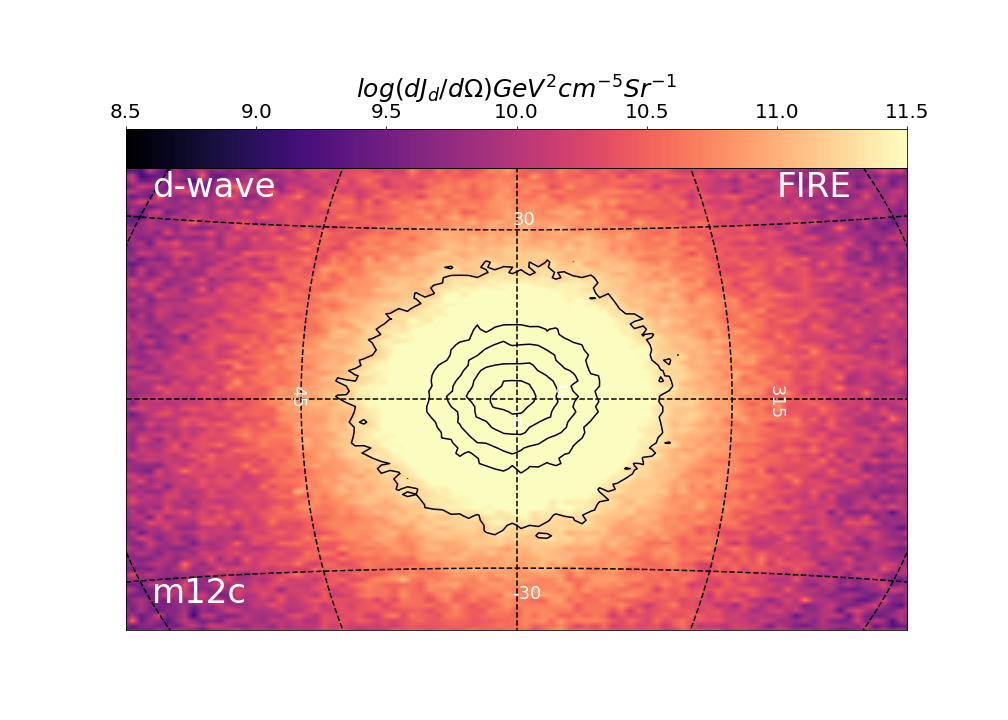}
    \caption{All sky maps of s-wave(top), p-wave(middle) and d-wave(bottom) J-factor for \texttt{m12c}.  The  DMO version is shown on the left and full physics FIRE on the right. The horizontal and vertical lines are spaced by 30 and 45 degrees respectively, to guide the eye. To emphasize the shape of the emission, black solid contours are drawn at a fixed fraction of the maximum pixel in each halo: 0.5, 0.2, 0.1, 0.05, and 0.01. Note that FIRE contours are more symmetric and circular than the DMO contours. Note the substructure in DMO maps. We see some mild flattening along the disk plane in the FIRE run.}
    \label{fig:map_m12c}
\end{figure*}

\subsection{Approach}
\label{sec:approach}

In what follows we aim to determine the astrophysical J-factors for each of our simulated halos for s-wave, p-wave, and d-wave annihilation.  In doing so we approximate the dark matter distribution $f(\vec{r}, \vec{v})$ as a separable function:
\begin{equation}
    f(\vec{r}, \vec{v}_1) = \rho(\vec{r}) \, g(\vec{v}(\vec{r})),
\end{equation}
with the dark matter density $\rho$ estimated using direct particle counts in the simulation. In this estimate, we use a cubic spline smoothing kernel \citep{Monaghan92} with smoothing length set to contain the mass of the nearest $32$ neighbors \citep[as described in][]{Hopkins15}.

For standard $s$-wave annihilation we have $Q(v)=1$ and the effective J-factor (Eq. \ref{eq:Jfactor_def}) reduces to a simple integral over the density squared:
\begin{eqnarray}
\frac{d J_{s}}{d \Omega} (\vec{\theta}) & = &
  \int \dd \ell \, \rho^2(\vec{r}) \int \dd^3 v_1 \, g_r(v_1)
  \int \dd^3 v_2 \, g_r(v_2)   \nonumber \\
   & = &  \int \dd \ell \, \rho^2[\ell(\vec{r})].
\end{eqnarray}
For $p$-wave annihilation, $Q(v) = (v/c)^2$, and Eq. \ref{eq:Jfactor_def} becomes

\begin{eqnarray}
\frac{d J_{p}}{d \Omega} (\vec{\theta}) & = &
  \int \dd \ell \, \rho^2(\vec{r}) \int \dd^3 v_1 \, g_r(v_1)
  \int \dd^3 v_2 \, g_r(v_2) \, \frac{|\vec{v}_1 - \vec{v_2}|^2}{c^2}  \nonumber \\
   & = & \frac{1}{c^2} \int \dd \ell \, \rho^2[\ell(\vec{r})] \, \mu_2(\ell(\vec{r})).
   \label{eqn:dp}
\end{eqnarray}

In the second line we have used $\mu_2$ to represent the second moment of the relative velocity at position $\vec{r}$.   
For $d$-wave annihilation, $Q(v) = (v/c)^4$, which implies

\begin{eqnarray}
\frac{d J_{d}}{d \Omega} (\vec{\theta}) & = &
  \int \dd \ell \, \rho^2(\vec{r}) \int \dd^3 v_1 \, g_r(v_1)
  \int \dd^3 v_2 \, g_r(v_2) \, \frac{|\vec{v}_1 - \vec{v_2}|^4}{c^4}  \nonumber \\
   & = & \frac{1}{c^4} \int \dd \ell \, \rho^2[\ell(\vec{r})] \, \mu_4(\ell(\vec{r})).
   \label{eqn:dd}
\end{eqnarray}

Here $\mu_{4}$ is the fourth moment of the relative velocity at position $\vec{r}$. We measure both $\mu_2$ and $\mu_{4}$ at each particle position using the nearest 32 dark matter particles.~ \footnote{For a perfectly spherically symmetric Maxwellian distribution, we expect the cross terms to vanish such that $\langle (\vec{v_1} - \vec{v_2})^2 \rangle  = 2 \sigma_v^2$ and $\langle (\vec{v_1} - \vec{v_2})^4 \rangle = 48 \sigma_v^4$ / 9, where $\sigma_{v}$ is the local velocity dispersion. We showed in \cite{2022MNRAS.513...55M} that direct measurement gives slightly lower estimates than would be expected from the simplified Maxwellian expectation.  We then construct all-sky maps of the relevant J-factors using appropriately-weighted and smoothed projections from mock observer locations (see \cite{2022MNRAS.513...55M}).}

\section{Methodology}

Previous studies have assumed a variety of shapes for the dark matter annihilation, informed by past numerical work on dark matter halo shapes  (e.g.  \cite{Abazajian20},\cite{zimmer2022andromeda}). There is a significant literature on the 3d shapes of dark matter halos, both in DMO simulations \cite{Allgood06} and with full hydrodynamics \cite{Chua22}.  Most of this theoretical work characterizes halos shapes by fitting the dark matter distributions to ellipsoidal configurations. These ellipsoidal fits must then be projected in the J-factor integral in order to determine the shape on the sky.  We chose to work directly with the most observationally-relevant predictions for indirect detection: the J-factor itself.

We specifically characterize the shape of the J-factor on the sky. See \cite{2022MNRAS.513...55M} for a detailed description of our calculation of these in simulations.  For our observing location, we choose the position in the simulation as viewed from the solar location, as described in \cite{2022MNRAS.513...55M}. 

Similarly to what was done in our previous work,\cite{2022MNRAS.513...55M} for FIRE halos, we start by rotating the disk of the galaxy by employing rotational matrices. Once we have rotated the disk, we choose a solar location 8.3 kpc from the galactic center as our observational point ( similar to what we would find from Gamma Ray telescope observations. For dark matter only runs, we choose an 8.3 kpc location that is arbitrary, since there is no disk. In both cases we simply use the x-axis that is defined in our new rotated frame of reference and thus the galaxy lies in the y-z plane. For DMO runs, there is no disk so we can simply choose this location along the x-axis for consistency. We then later convert to angular coordiantes for the skymaps and analysis.

We find that, at a fixed $dJ/d\Omega$ contour, the shape is well-fit by an ellipse, with semi-major axis $R_{\rm major}$ and semi-minor axis $R_{\rm minor}$ measured in degrees on the sky. We use least squares fitting to fit ellipses to the J-factor emission. The ratio of minor to major axes ($R_{\rm minor}/R_{\rm major}$) allows us to characterize the shape of each contour, with ratio $1.0$ corresponding to circular/spherical emission.

We provide fits to contours at specific J-factor values for s,p,d wave annihilation models and fits to contours a specific fraction of peak emission for each halo.  This provides a way to measure both ``brightness" or ``size" of emission (at fixed J, larger $R_{\rm major}$ values are brighter and bigger on the sky) and also the shape or ``flatness" of emission on the sky (at fixed fraction of peak J, larger values of $R_{\rm major}$ have flatter emission profiles).Our results are illustrated in the figures to follow and summarized in Tables  \ref{tab:one} -- \ref{tab:two}. The first table (\ref{tab:one}) presents $R_{\rm major}$ and $R_{\rm minor}$ values and associated axis ratios for each of our halos (DMO and FIRE) measured at fixed physical values of d$J/$d$\Omega$ for both unpaired and paired runs,and for s, p, and d-wave dark matter. The second table (\ref{tab:two}) shows the same information for  d$J/$d$\Omega$ contours measured at $20\%$ of the peak flux for each halo. The second table also provides the orientation angle of $R_{\rm major}$ with respect to the Galactic plane in each FIRE simulation.

\subsection{Geometric Setup}

For each halo in our sample, we calculate J-factors as defined in Equation \ref{eq:Jfactor_def}, integrating from a mock Solar location (setting $\ell = 0$) to the edge of the halo, which we define as a sphere of radius $r = 300$ kpc from the center of each halo in every case.  While the virial radii \citep{ByranNorman1998} of our halos range from $300 - 335$ kpc, we fix $300$ kpc as the halo boundary for consistency.  Since most of the J-factor signal comes from the inner halo, changing the outer radius by $10\%$ has no noticeable affect on our results. 

For the DMO runs, we assume that the Galactic Center corresponds to the halo center and fix the observer location to be at a distance 8.3 kpc from the halo center along the x-axis of the simulation.  For FIRE runs, we position the observer in the galaxy disk plane at a radius of 8.3 kpc from the halo center.  We define the disk plane to be perpendicular to the angular momentum vector of all the stars within 20 kpc of the central galaxy.

\begin{table*}
	\centering
	\caption{ Ellipse fits characterizing the shape of on-sky s-wave dJ/d$\Omega$ maps for each m12 simulation.  Fits are to contours at 20 percent of the peak dJ/d$\Omega$ value in the map.(1) Simulation name (2) The semi-major axis of the elliptical fit (3) The semi-minor axis of the elliptical fit (4) The ratio of minor-to-major axis of the elliptical fit. (5) Orientation angle of major axis with respect to the Galactic plane. Note that the angle for DMO is not given since the orientation is arbitrary in that case. (6)-(9) same for p-wave (10)-(13) same for d-wave
	   }
        \label{tab:two}
	\begin{tabular}{lcccc @{\spacer} cccc @{\spacer} cccc } 
		\hline
		Simulation & major axis & minor axis & ratio & angle &  major axis & minor axis & ratio & angle &  major axis & minor axis & ratio & angle     \\ 
		           &   &  s-wave &  & 
		             &   & p-wave   & &   &   & d-wave &    \\

            \hline
		\mi       &  $12.5^{ \circ }$ &  $10.1^{ \circ }$   &  0.810 &  $2.8^{ \circ }$   &   $10.7^{ \circ }$     &  $8.4^{ \circ }$  & 0.78  & $2.5^{ \circ }$  &   $9.32^{ \circ }$ &  $7.30^{ \circ }$  & 0.781 & $0.87^{ \circ }$      \\
		
        \midark      &  $11.8^{ \circ }$  & $6.20^{ \circ }$ & 0.520 & -  & $25.2^{ \circ }$ & $13.4^{ \circ }$ &  0.53 &-    &  $26.7^{ \circ }$  & $14.3^{ \circ }$ & 0.533 &-     \\ \\
		
		\mc         &  $14.1^{ \circ }$  &   $9.92^{ \circ }$   & 0.701   & $0.66^{ \circ }$     &  $11.6^{ \circ }$   &  $8.45^{ \circ }$   & 0.730   &  $0.98^{ \circ }$       &    $10.0^{ \circ }$  &   $6.90^{ \circ }$  &  0.682  & $1.2^{ \circ }$     \\

  \mcdark      &  $11.6^{ \circ }$ &  $5.51^{ \circ }$ & 0.480  & -  &  $16.7^{ \circ }$  & $8.92^{ \circ }$  & 0.530      &  -   &   $20.0^{ \circ }$ & $10.6^{ \circ }$  & 0.531   & -   \\ \\

  \mm         & $19.6^{ \circ }$ & $12.7^{ \circ }$  & $0.650$ &  $1.2^{ \circ }$ &   $16.48^{ \circ }$  & $11.2^{ \circ }$  & 0.68 & $2.8^{ \circ }$  & $14.4^{ \circ }$ & $9.71^{ \circ }$   & 0.670  & $1.5^{ \circ }$     \\
		
  \mmdark     &  $9.50^{ \circ }$   &  $6.51^{ \circ }$ & 0.690  & - &  $13.5^{ \circ }$ & $10.1^{ \circ }$ & 0.750 & - & $17.4^{ \circ }$  & $13.2^{ \circ }$ & 0.760 & -   \\ \\
		
		\mf          &  $14.6^{ \circ }$   & $11.5^{ \circ }$ & 0.784 & $1.0^{ \circ }$  &    $12.3^{ \circ }$   &  $9.48^{ \circ }$  & 0.774 &  $3.1^{ \circ }$       & $10.5^{ \circ }$   & $8.08^{ \circ }$ & 0.770 & $4.5^{ \circ }$         \\
		
        \mfdark     &  $12.5^{ \circ }$   &  $6.12^{ \circ }$  & 0.490 & -  & $15.2^{ \circ }$   & $7.43^{ \circ }$ & 0.490  & - & $15.5^{ \circ }$   & $9.72^{ \circ }$  & 0.627  & -     \\ \\
		
		\mw        &  $9.54^{ \circ }$   &  $7.06^{ \circ }$   & 0.740  & $2.0^{ \circ }$   &  $8.67^{ \circ }$ &   $6.19^{ \circ }$ & 0.714 &  $4.0^{ \circ }$    &  $7.78^{ \circ }$  & $5.76^{ \circ }$ & 0.740 & $4.0^{ \circ }$   \\
		\mwdark     &  $8.73^{ \circ }$  &  $6.02^{ \circ }$   &  0.690 &  -    & $6.13^{ \circ }$   & $4.49^{ \circ }$ & 0.733 & -    &  $25.8^{ \circ }$ & $16.8^{ \circ }$ & 0.652 & -  \\ \\
		
		\mb          &  $14.1^{ \circ }$  &  $11.8^{ \circ }$  & 0.842  & $2.4^{ \circ }$   &  $11.36^{ \circ }$  &  $9.33^{ \circ }$ & 0.823 &  $1.0^{ \circ }$       &   $9.44^{ \circ }$ & $7.86^{ \circ }$  & 0.832 &  $7.0^{ \circ }$  \\
		\mbdark        & $6.00^{ \circ }$ & $5.64^{ \circ }$ & 0.943 & -  & $9.64^{ \circ }$ &  $9.17^{ \circ }$ & 0.950  & -   & $13.3^{ \circ }$  & $12.4^{ \circ }$ & 0.933  &-     \\ \\

		\Romeo       &  $8.90^{ \circ }$  & $6.57^{ \circ }$ & 0.740  & $0.84^{ \circ }$  &  $8.66^{ \circ }$ &  $6.29^{ \circ }$  & 0.726 & $1.23^{ \circ }$        &  $8.37^{ \circ }$ & $6.05^{ \circ }$  & 0.723 & $1.9^{ \circ }$     \\
		
        \Romeodark   &  $5.36^{ \circ }$   &  $4.35^{ \circ }$ & 0.810 & -   &  $7.73^{ \circ }$  & $6.47^{ \circ }$  & 0.837   & -   &    $11.2^{ \circ }$  &  $9.548^{ \circ }$  & 0.853 & -     \\ \\
	
			\Juliet     &   $8.24^{ \circ }$   &  $7.01^{ \circ }$ & 0.850 & $1.3^{ \circ } $   &  $7.32^{ \circ }$  & $6.2^{ \circ }$  & 0.86 & $1.84^{ \circ }$    &  $6.6^{ \circ }$   & $5.40^{ \circ }$  & 0.82 & $3.0^{ \circ }$   \\
		\Julietdark  &   $5.72^{ \circ }$  & $4.20^{ \circ }$  & 0.740  &-   &   $9.31^{ \circ }$  & $6.51^{ \circ }$  & 0.701  &-   & $13.3^{ \circ }$  & $9.4^{ \circ }$ & 0.710 & -    \\ \\

		\Thelma      &  $20.5^{ \circ }$ &  $15.7^{ \circ }$ & 0.765   & $0.70^{ \circ }$  &  $17.8^{ \circ }$  & $13.7^{ \circ }$  & 0.768 & $1.8^{ \circ }$ &   $15.5^{ \circ }$ & $12.2^{ \circ }$  & 0.786 & $4.1^{ \circ }$  \\
		\Thelmadark   &  $11.6^{ \circ }$  & $4.81^{ \circ }$   & 0.415 & -    &  $18.8^{ \circ }$   & $8.47^{ \circ }$   & 0.449 & -   &  $26.9^{ \circ }$  &  $12.7^{ \circ }$ & 0.471 & -     \\  \\

		\Louise      &  $12.9^{ \circ }$  & $10.1^{ \circ }$   & 0.780  & $1.0^{ \circ }$ &   $11.9^{ \circ }$  & $9.21^{ \circ }$  & 0.772 & $1.7^{ \circ }$     &  $10.7^{ \circ }$  & $8.60^{ \circ }$  & 0.804 & $0.37^{ \circ }$   \\
		\Louisedark   &  $6.96^{ \circ }$  & $5.87^{ \circ }$   & 0.844  & -  &  $13.1^{ \circ }$ & $10.2^{ \circ }$  & 0.773 & -  & $17.0^{ \circ }$ & $13.6^{ \circ }$ &  0.803 & -   \\  \\

		\Romulus     &  $5.84^{ \circ }$  & $4.57^{ \circ }$  & 0.783 & $1.0^{ \circ }$    &  $5.42^{ \circ }$ &  $4.30^{ \circ }$  & 0.793 &  $1.0^{ \circ }$   &  $5.10^{ \circ }$ & $4.01^{ \circ }$ & 0.786  & $1.0^{ \circ }$  \\

    \Romulusdark  &  $10.2^{ \circ }$  & $5.76^{ \circ }$ & $0.565$ & -    &  $19.2^{ \circ }$  & $10.1^{ \circ }$  & $0.527$ & -   &  $25.7^{ \circ }$  & $15.2^{ \circ } $ & 0.590   & -   \\ \\

		\Remus    &  $9.67^{ \circ }$  &  $7.47^{ \circ }$  & 0.772 & $1.1^{ \circ }$  &  $8.96^{ \circ }$    & $6.77^{ \circ }$  & 0.755 & $1.3^{ \circ }$   &  $8.27^{ \circ }$  & $6.18^{ \circ }$ & 0.748 & $1.6^{ \circ }$ \\

       \Remusdark  &  $8.04^{ \circ }$ & $5.25^{ \circ }$  & 0.652 & -  &  $11.7^{ \circ }$ &  $7.94^{ \circ }$ & 0.681 & -    &   $15.1^{ \circ }$ &  $10.5^{ \circ }$  & 0.696 & -  \\ \\

		\hline
	\end{tabular}
\end{table*}



\section{J-factor Emission Sky Maps}

Figures 1 and 2 show example all-sky J-factor maps for \texttt{Juliet}, \texttt{JulietDMO}, \texttt{M12c}, and \texttt{M12cDMO} for s-wave, p-wave, and d-wave cases. For each of the FIRE runs, the disk axis is oriented horizontally.

In each figure, the DMO version is shown on the left, while the FIRE version is shown on the right.  The color bars are the same for each pair of halos, though they change depending on whether it is s-wave, p-wave, or d-wave dark matter. The black lines, on the other hand, are contours plotted at a constant peak emission fraction for each map specifically. The lines correspond to 50,20,10,5, and 1 percent of the peak emission level, giving us 5 contours.  Here we summarise the results for each type of dark matter (s-wave, p-wave, and d-wave).

This is due to our choice of aligning our observation with the disk in this way. We have specifically chosen each galaxy center to be oriented with the disk horizontal to our observation. In practice this may not be possible, and depends on the orientation of the galaxy with respect to our observation. However, with simulations this is always possible. The orientation of the DMO halo is arbitrary, given that there is no disk, and only dark matter particles. In the figures 1 and 2, the dotted horizontal and vertical lines are spaced by 30 and 45 degrees respectively, to guide the eye. These are standard angluar coordinates. 

The intensity lines of the J-factor signal show how the intensity of the emission signal varies with angle as measured outward from the center. Note that the emission signal intensity is far more spherical in the case of FIRE runs than it is for Dark Matter Only runs. In DMO runs the signal is elongated along the polar angle axis in every direction. Meanwhile, the FIRE runs are always centralized and very symmetric. What this tells us is that we should expect to see a roughly spherically symmetrical signal when looking at the galactic center excess in just the way that we have chosen (with the disk axis oriented horizontally. Since we live in a baryonic universe, there are interesting implications for a possible signal. What this tells us is that the galactic signal should be symmetric when viewed horizonal, and should not vary much when we are at the same radial distance from the center, regardless of which side we look. On the other hand, the dark matter only signal is quite jagged and irregular and varies greatly even at the same radial distance. When looking at future data from Gamma Ray Telescopes, it will be important to determine whether this relatively constant signal with radius from the galactic center is actually observed. If so, this would be consistent with what we have found. 

We will now give a brief summary of the s,p,and d wave contour maps and their implications for indirect detection. 

\subsection{s-wave maps}
In both s-wave maps (Figures 1 and 2)  $\bf{top}$, substructure is more apparent in DMO and much less so in FIRE. We see that the contours are more elongated in DMO runs compared to FIRE. Interestingly, the faintest contours are slightly asymmetrical for DMO, with centers that do not seem to be in alignment (which is to say, the centers of the innermost contour don't match with the outermost). Meanwhile, FIRE contours have centers which appear well-aligned with the center of the galaxy as defined by our halo finder.  The contours for FIRE are more circular. The contour for the 50 percent peak emission is generally larger for FIRE, as expected for an emission profile that is a little bit flatter in the center.  While none of the emission contours perfectly "centered", they are consistent with zero within our numerical resolution, which is roughly 2.7 degrees.

Note that the contours are oriented vertically with respect to the Galactic plane in \texttt{m12cDMO}.  However, the orientation of the plane was chosen at random in this DMO case since there is no disk in the simulation. The FIRE halos, of course, do have disks, so the Galactic plane in those cases are physically relevant.  Interestingly, both \texttt{m12c} and \texttt{Juliet} we see some mild flattening of the contours along the disk planes.  We find that this is generally the case for all of the FIRE halos we examine: the flattening is always along the disk plane. Implications for these findings are that we would expect a signal range at the very center to be approximately $10^{24} (Gev)^{2}cm^{-5}Sr^{-1}$ at the very center of the galaxy if the source of our signal is s-wave annihilation, with a drop of 99 percent when we have moved to approximately within 35-45 degrees outside the center. This should serve as a reasonable target for future surveys of the galactic center excess.

\subsection{p-wave maps}

 The $\bf{middle}$ panels of Figures 1 and 2 show substructures are much less apparent for p-wave DMO maps compared to their s-wave counterparts. The DMO contours appear are quite stretched out compared to the FIRE contours of p-wave maps, and also much more stretched out and elongated than contours of DMO for s-wave cases. This is because, relative to their peak central emission, p-wave J-factor profiles are flatter than their s-wave counterparts in the DMO case (as can be seen in Figures 1 and 2) $\bf{middle}$. Note that the actual axis ratios of the DMO contours are similar to those of s-wave case. The FIRE contours are quite similar between s-wave and p-wave. The FIRE p-wave contours resemble those of the s-wave contours, in that they are flattened mildly along the Galactic plane.

 Implications for these findings are that we would expect a signal range at the very center to be approximately $10^{17.5}(Gev)^{2}cm^{-5}Sr^{-1} $ at the very center of the galaxy if the source of our signal is p-wave annihilation, with a drop of 99 percent when we have moved to approximately within 30-40 degrees outside the center.

\subsection{d-wave maps}

The $ \bf{bottom}$ panels of Figures 1 and 2 show the d-wave sky maps for \texttt{Juliet} and \texttt{m12c}. As in the p-wave cases, the DMO d-wave maps, and associated contours, have significantly different shapes than in the s-wave case. The FIRE runs, on the other hand, are much more similar, with less variation for the assumed mode of annihilation. The FIRE contours are much rounder, and reveal a more centrally-concentrated emission profile. Furthermore, d-wave models and p-wave models are more centrally concentrated than s-wave, meaning the fall of in intensity is sharper for p and d wave.

Implications for these findings are that we would expect a signal range at the very center to be approximately $10^{11.8} (Gev)^{2}cm^{-5}Sr^{-1}$ at the very center of the galaxy if the source of our signal is d-wave annihilation, with a drop of 99 percent when we have moved to approximately within 25-35 degrees outside the center.

\section{Population Results: Elliptical Fits on the Sky}
\label{sec:sample}
Tables 1 and 2 provide a summary of shape information of on-sky ellipse fits for all of our halos, listed in DMO and FIRE pairs in each case. 
In Table 1, we list $R_{\rm major}$ and $R_{\rm minor}$ values and associated axis ratios fit to contours set at fixed dJ/d$\Omega$ values that are representative of the bright inner regions of halos for s-, p-, and d-wave models. These tables also list the ratio of minor to major axis, which provides a measure for how far the emission deviates from circular (a value of 1.0 corresponds to perfect circle).  Our fiducial choices for s-wave, p-wave, and d-wave contours are $10^{24} (Gev)^{2}cm^{-5}Sr^{-1}$, $10^{17.5}(Gev)^{2}cm^{-5}Sr^{-1} $, and $10^{11.8} (Gev)^{2}cm^{-5}Sr^{-1}$, respectively. Though, for a subset of the dimmest DMO/FIRE halo pairs, we were forced to chose slightly smaller values for the d$J$/d$\Omega$ contours in order to achieve a smooth fit (see Table captions). By measuring the shapes and angular extent of contours at a fixed dJ/d$\Omega$ value, we are able to compare and contrast the ``brightness" or ``size" of emission: at fixed values, larger $R_{\rm major}$ values are brighter and bigger on the sky.

\begin{figure*}
			\includegraphics[height =0.6\columnwidth,trim = 0 0 0 0]{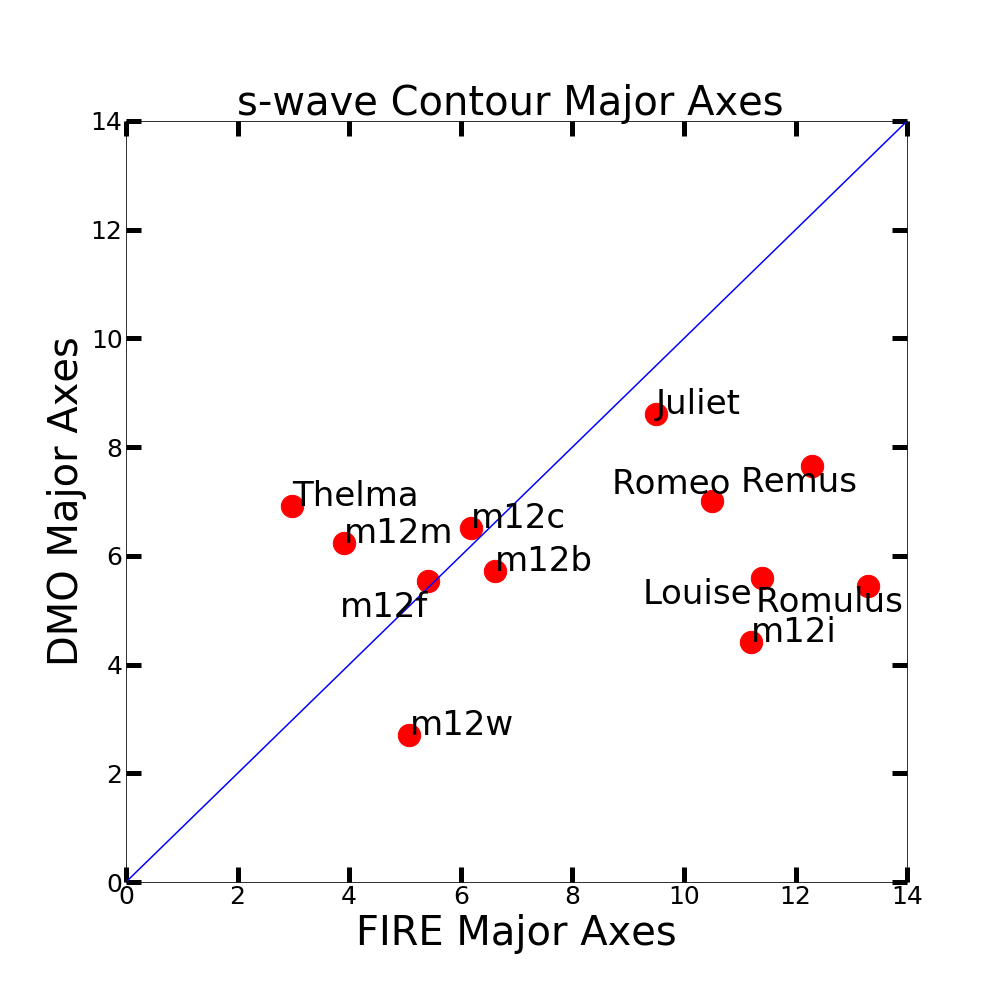} \hspace{.1in}
		\includegraphics[height =0.6\columnwidth, trim = 0 0 0 0]{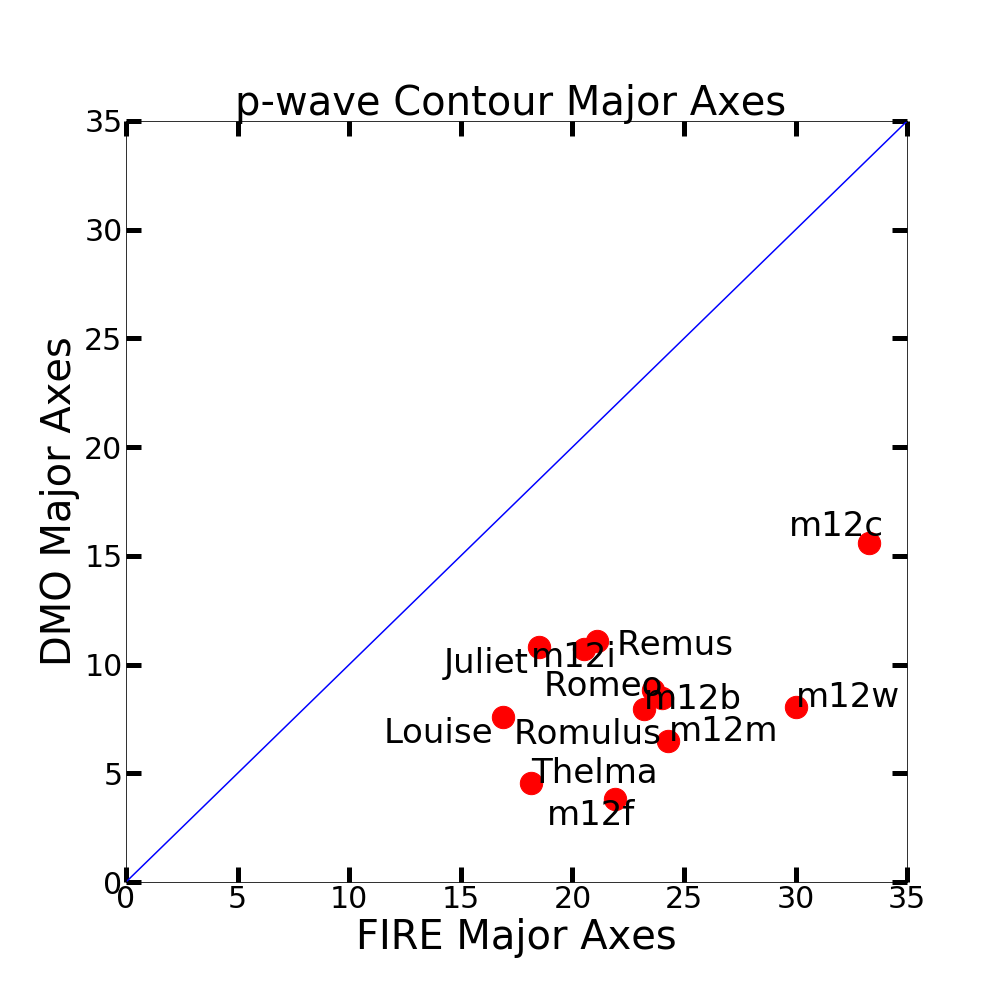} \hspace{.1in}
			\includegraphics[height =0.6\columnwidth, trim = 0 0 0 0]{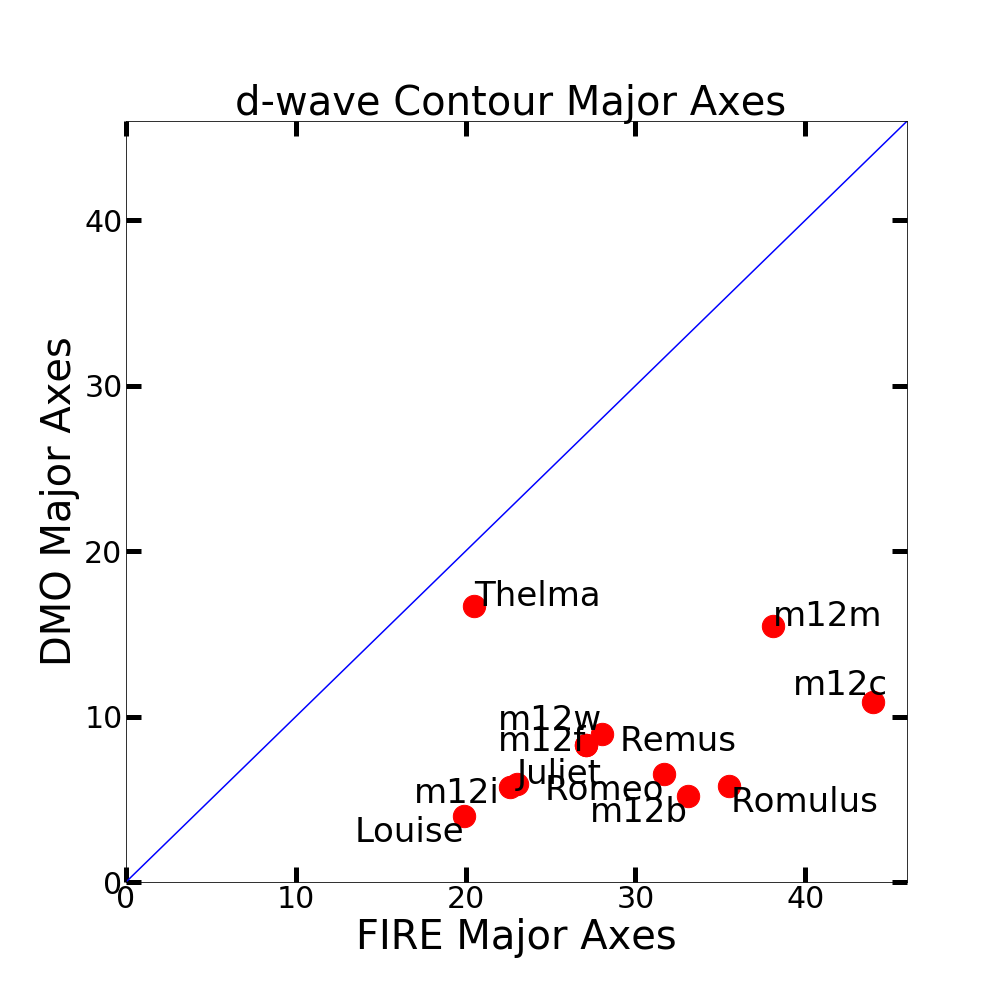}
    \caption{ {\bf left }: Best-fit ellipse-fitted major axis values in degrees for DMO runs (vertical)
and FIRE runs (horizontal). These ellipse fits correspond to a fixed contour value of $ 10^{24} (Gev)^{2}cm^{-5}Sr^{-1}$ on the sky. {\bf center }: Same for p-wave. These ellipse fits correspond to contours at $10^{17.5} (Gev)^{2}cm^{-5}Sr^{-1}$. {\bf right }: same for d-wave. These ellipse fits correspond to contours at $10^{11.8} (Gev)^{2}cm^{-5}Sr^{-1}$.
 }
    \label{fig:contour_major_axes}
\end{figure*}

\begin{figure*}
			\includegraphics[height =0.6\columnwidth,trim = 0 0 0 0]{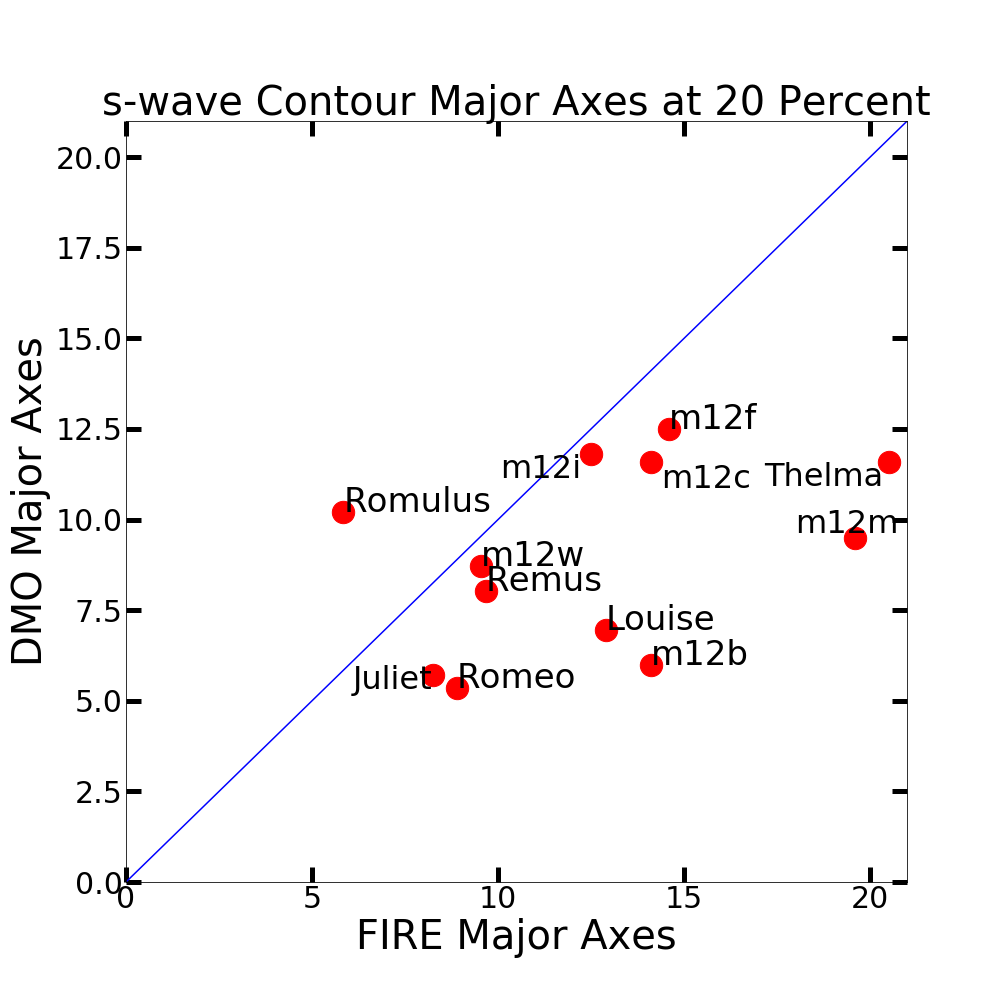} \hspace{.1in}
		\includegraphics[height =0.6\columnwidth, trim = 0 0 0 0]{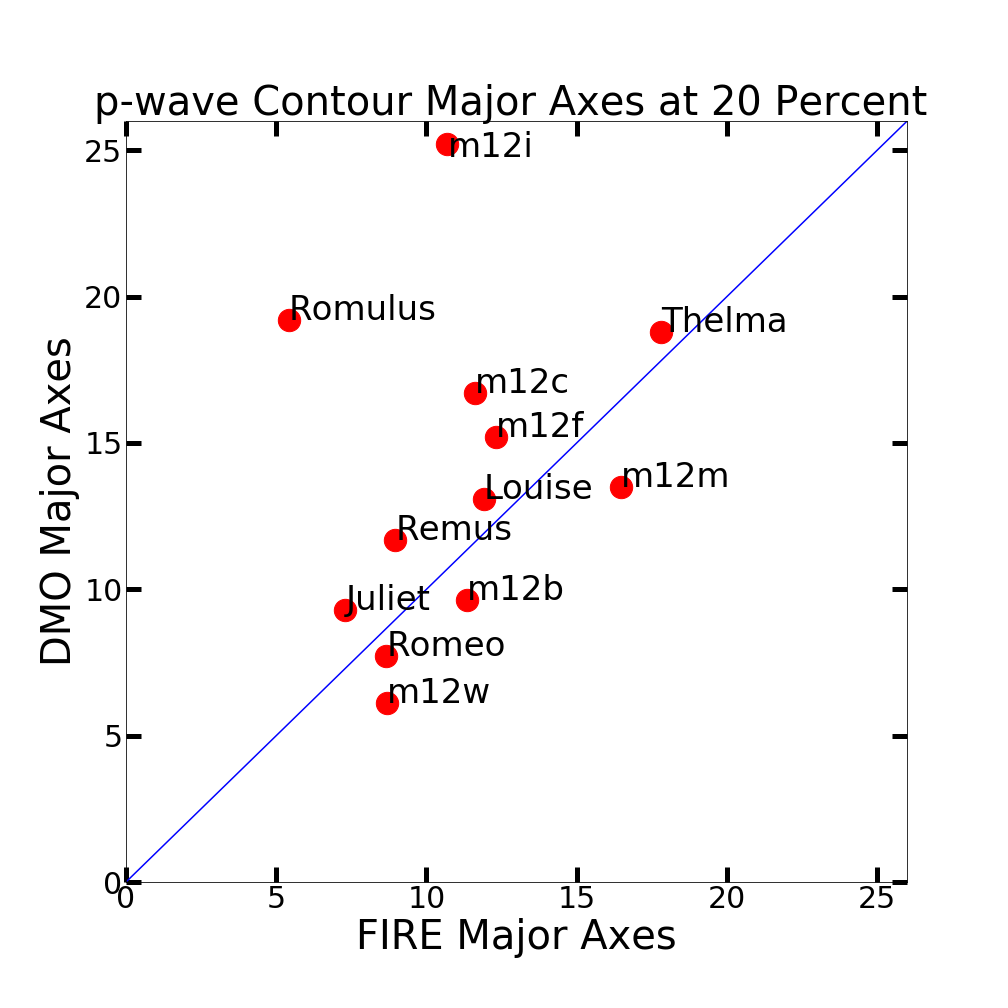} \hspace{.1in}
			\includegraphics[height =0.6\columnwidth, trim = 0 0 0 0]{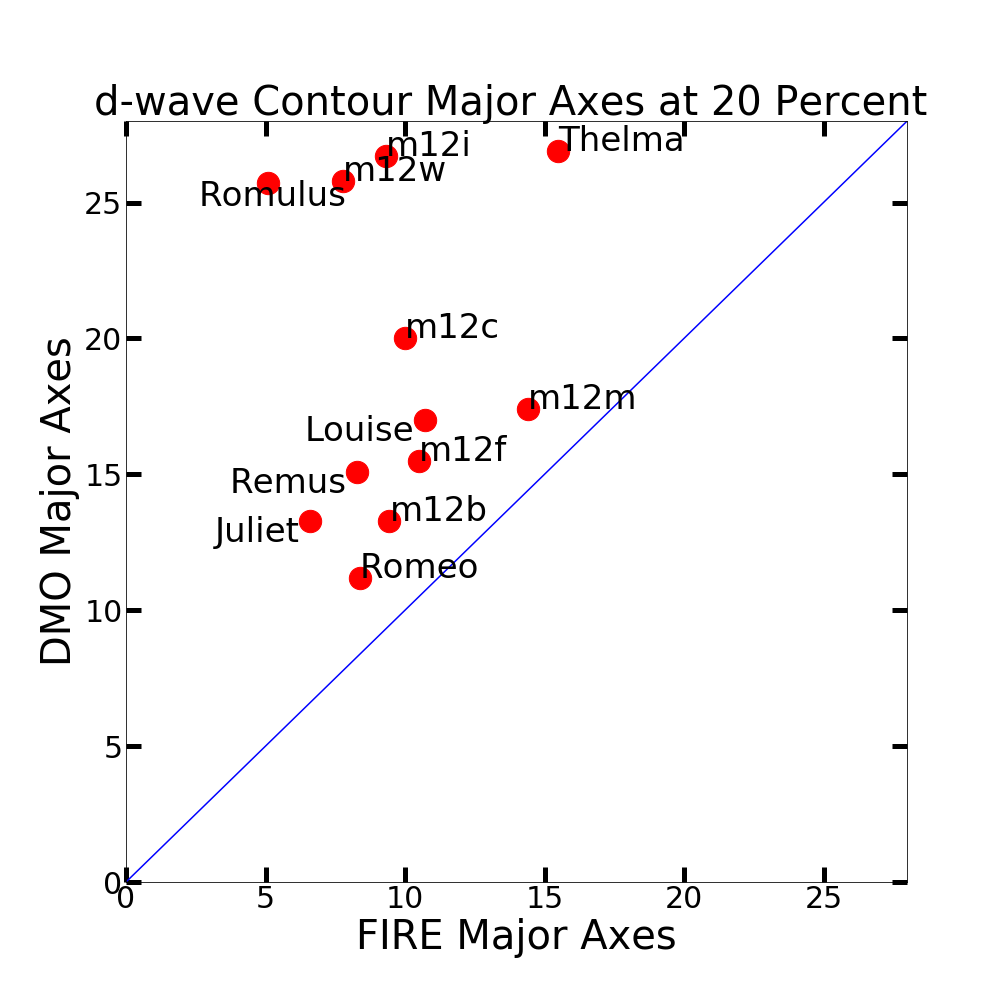}
    \caption{Best-fit ellipse-fitted major axis values in degrees for DMO runs (vertical) and FIRE runs (horizontal) at 20 percent the maximum value in the map.   {\bf left }: We see that the FIRE runs are typically more extended at fixed fraction of peak emission, meaning that the profiles are ``flatter" in shape.  {\bf middle }: for p-wave, ellipse fits to contours at $10^{17.5} (Gev)^{2}cm^{-5}Sr^{-1}$.  {\bf right }:for d-wave, ellipse fits to contours at $10^{11.8} (Gev)^{2}cm^{-5}Sr^{-1}$ }
    \label{fig: contour_major_axes_20_percent}
\end{figure*}


\begin{figure*}
		\includegraphics[height =0.6\columnwidth, trim = 0 0 0 0 ]{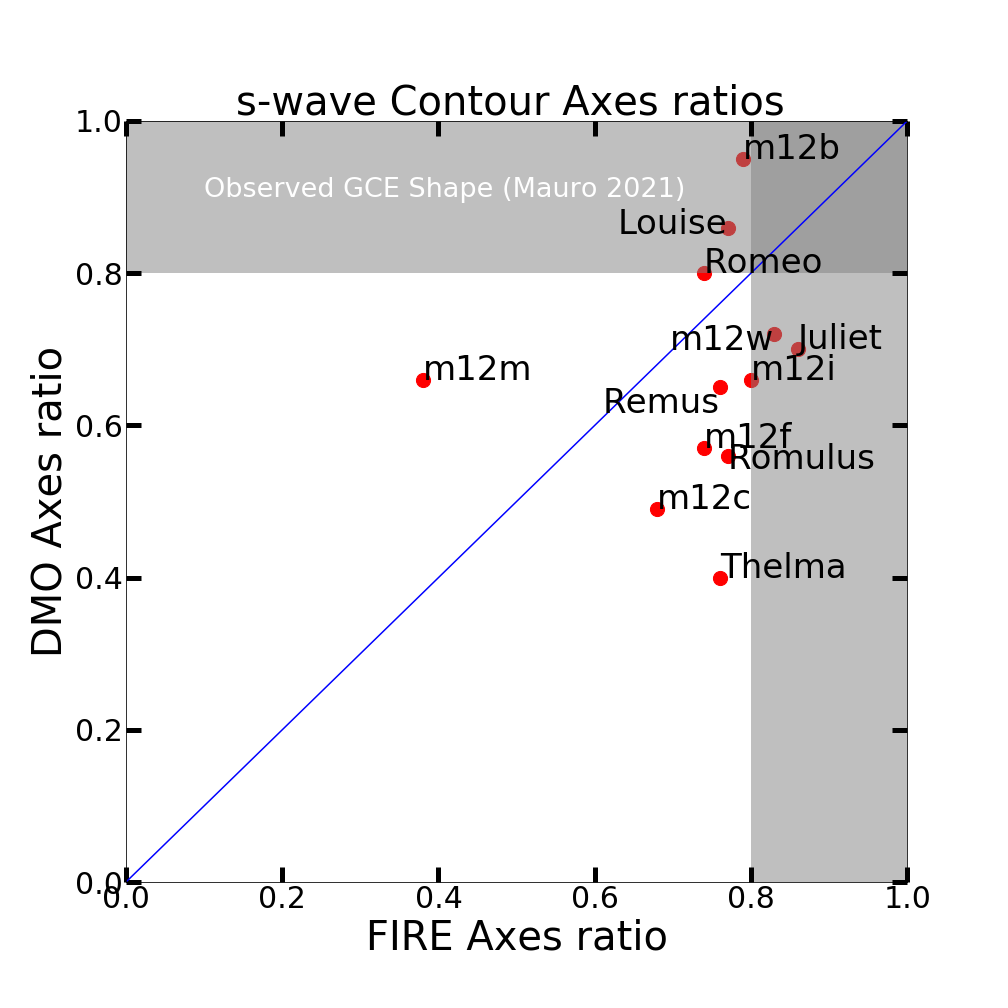} \hspace{.1in}
				\includegraphics[height =0.6\columnwidth, trim = 0 0 0 0 ]{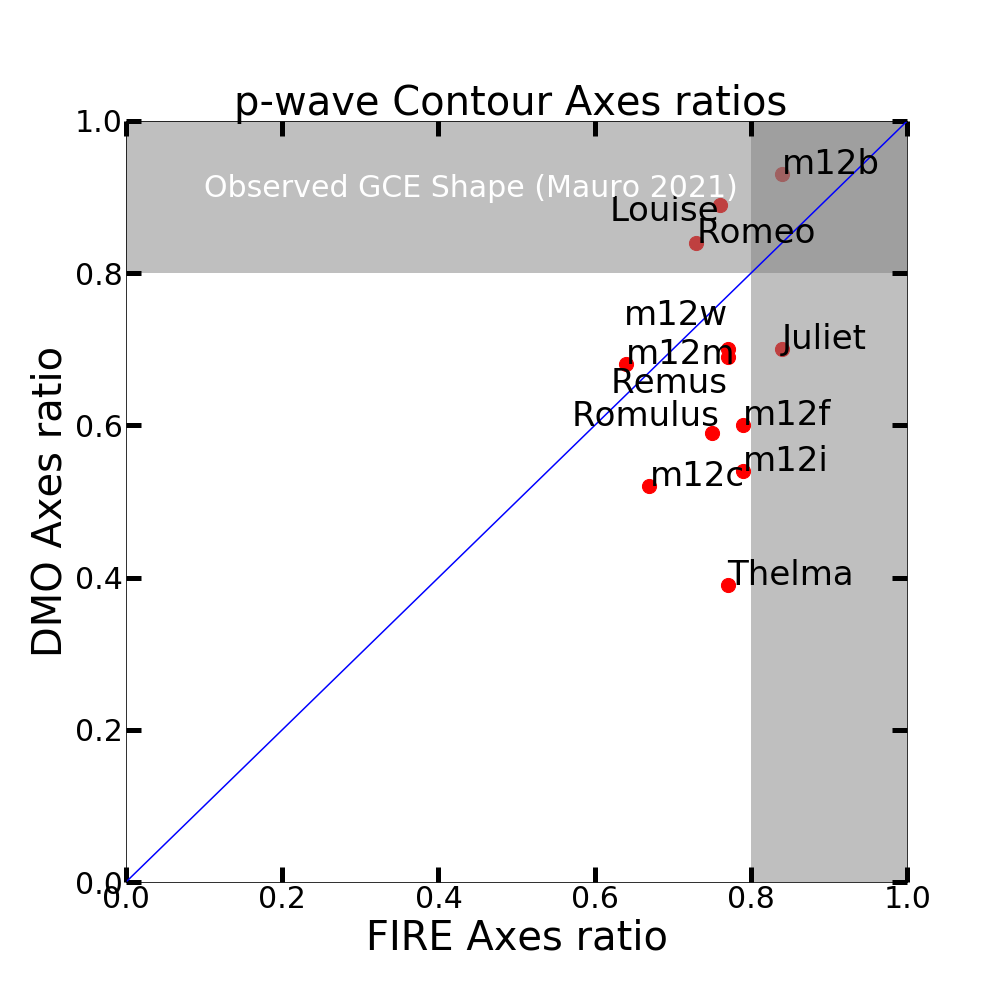}\hspace{.1in}
						\includegraphics[height =0.6\columnwidth, trim = 0 0 0 0 ]{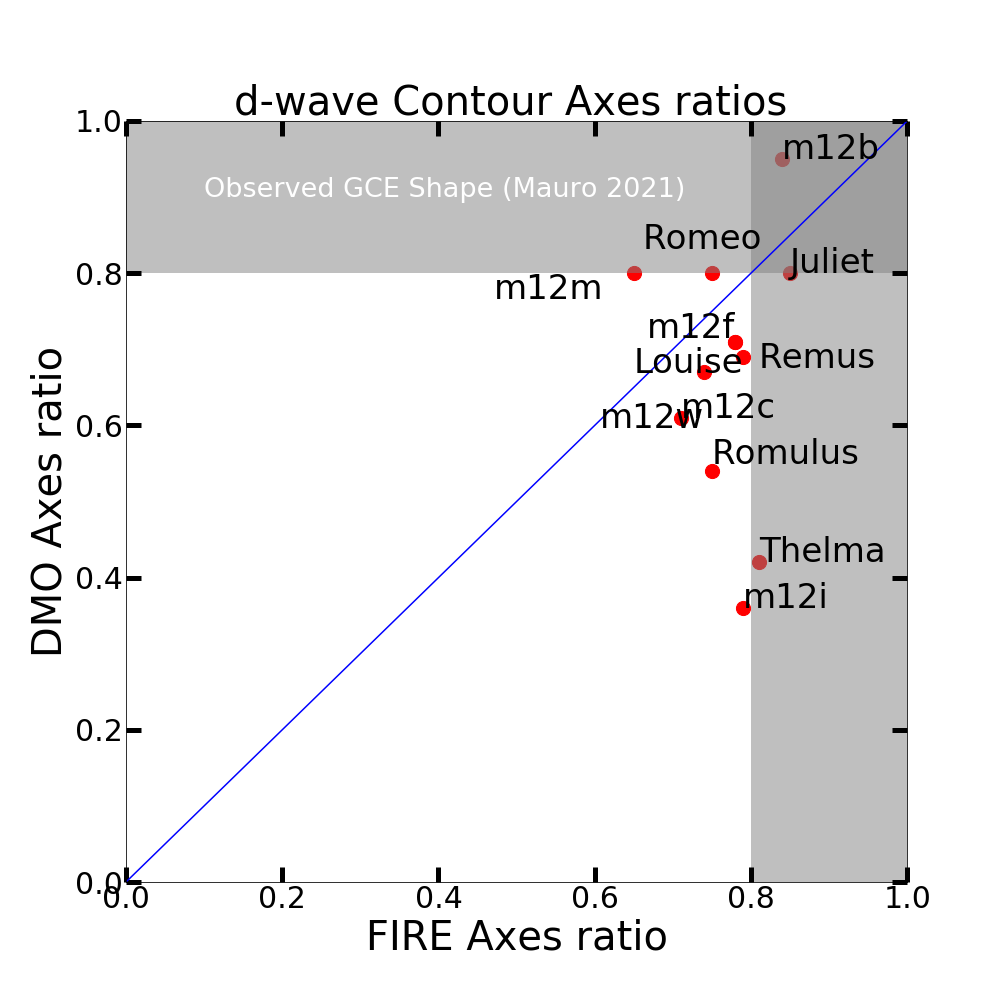}
    \caption{ {\bf left : } Ratios of axes (semi-minor/semi-major) of contour fits for s-wave for fixed value DJ/DO ($10^{24}$) Note that for s-wave the values are less constrained than for ratios taken at 20 percent values (see Figure 6) and vary from 0.65 to 0.85 for FIRE, while for DMO the
axes range from below 0.4 to almost 1. {\bf middle :}  contour fits for p-wave for fixed value DJ/DO ($10^{17.5}$) {\bf right:} contour fits for d-wave for fixed value DJ/DO ($10^{11.5}$) }   
    \label{fig:contour_axes_ratios}
\end{figure*}

\begin{figure*}
			\includegraphics[height =0.6\columnwidth,trim = 0 0 0 0]{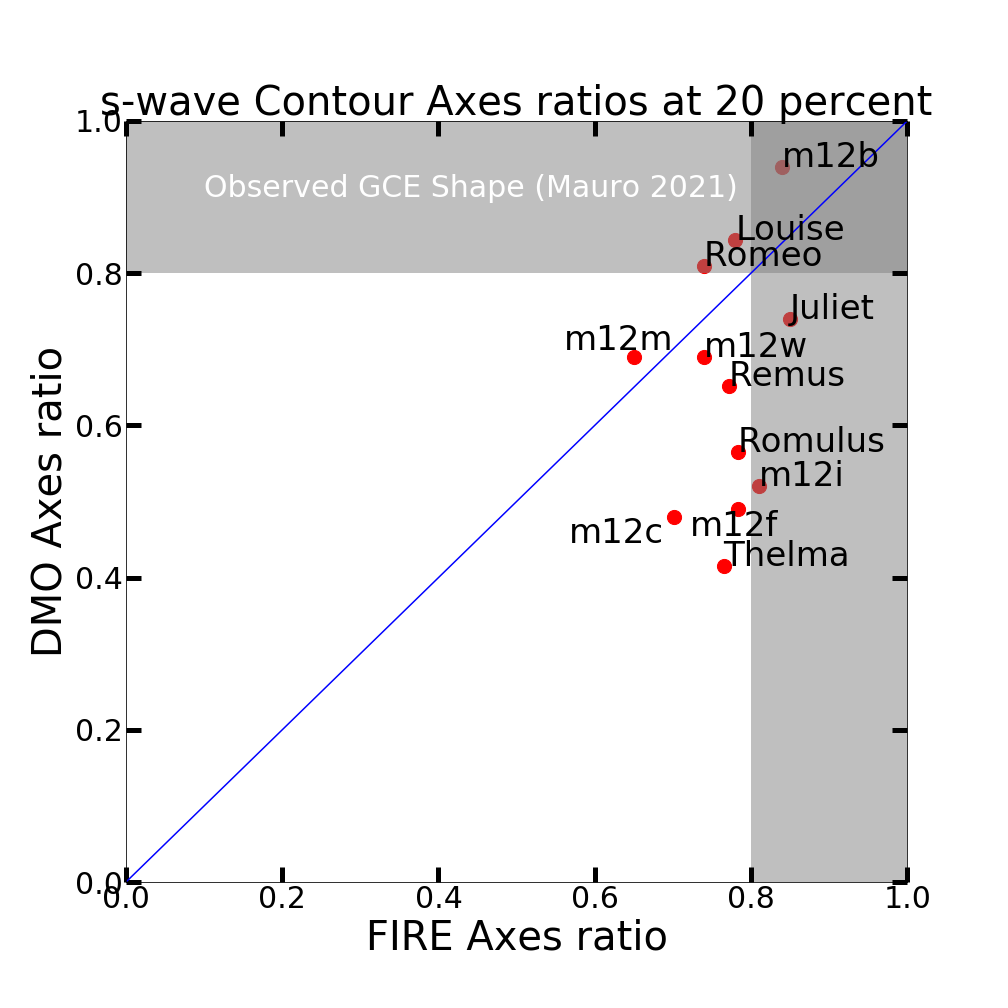} \hspace{.1in}
		\includegraphics[height =0.6\columnwidth, trim = 0 0 0 0]{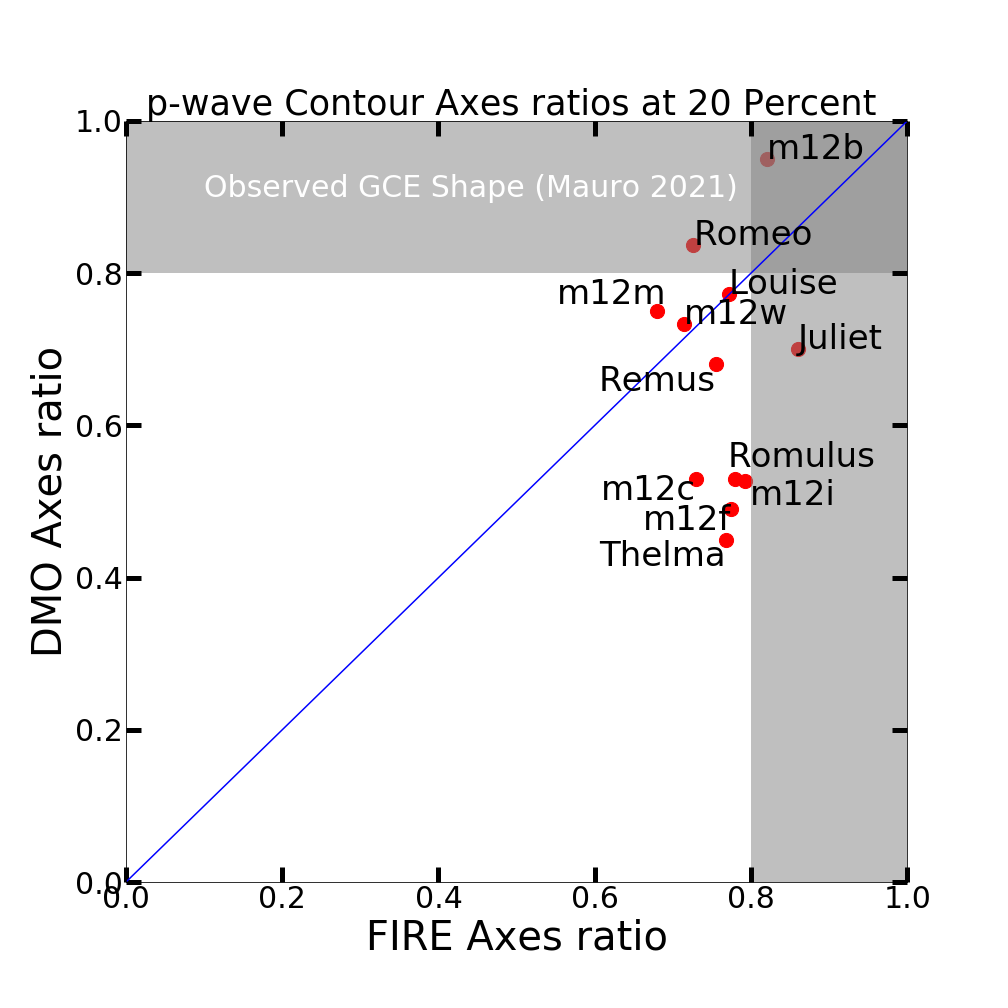} \hspace{.1in}
			\includegraphics[height =0.6\columnwidth, trim = 0 0 0 0]{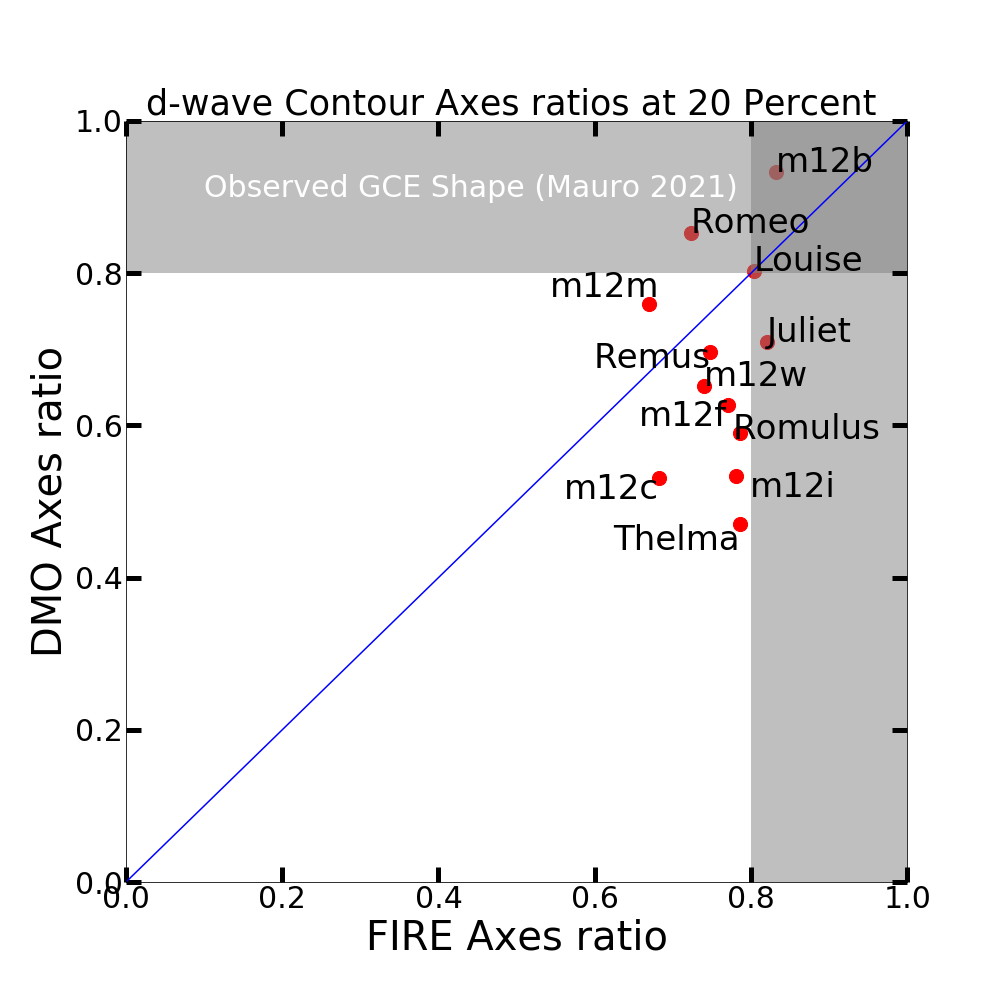}
    \caption{ {\bf left }Ratios of axes (semi-minor/semi-major) of contour fits for s-wave at the 20 percent peak value of DJ/DO for the given halo. {\bf middle } Ratios of axes (semi-minor/semi-major) of contour fits for p-wave at the 20 percent peak value of DJ/DO for the given halo. Note that FIRE halos have ratios of semi-minor to semi-major that are close to 0.8, while for DMO the axes range 0.4 to almost
1. {\bf right } Ratios of axes (semi-minor/semi-major) of contour fits for d-wave at the 20 percent peak value of DJ/DO for the given halo. Note that for s,p, and d-wave FIRE the ratios of semi-minor to semi-major are even more constrained and even closer to the value of 0.8, (d-wave fire in particular) while for DMO the axes range slightly above 0.4 to almost 1 }
    \label{fig:contour_axes_ratios_20_percent}
\end{figure*}

Table 2 provides similar information, but now using dJ/d$\Omega$ contours measured at $20\%$ of the peak flux for each halo individually. We choose 20 percent of the peak flux because this provides a consistent baseline to compare across different halos. Since the halo size of each run varies significantly, this provides a quantitative and consistent way to study the shape and angular extent of the emission signal. By measuring the shapes and angular extent of contours at a fixed fraction of peak dJ/d$\Omega$ emission, we provide information on the relative ``flatness" of emission on the sky: larger values of $R_{\rm major, 20}$ have flatter emission profiles, smaller $R_{20}$ values are more peaked. For the FIRE versions of each halo, we also list the orientation angle of the major axis fit with respect to the Galactic plane.  Recall that we have oriented our observer in the plane of the disk formed in each FIRE simulation, so the orientation angle is meaningful. The DMO versions are observed from an arbitrary disk plane, so the orientation with respect to the galactic plane is not physically meaningful.

As expected from the previous figures, we see from Table 2 that the major axes of the contours are always aligned to within 4 degrees of the plane, with one outlier (\texttt{M12b} d-wave) aligned at 7 degrees.  This level of alignment provides a potentially powerful prior for the expected alignment of emission signals from dark matter annihilation.

Figures 3 and 4 show the best-fit major-axis values for DMO runs (vertical axis) versus FIRE runs (horizontal axis). In each case in Figure 3 the major axis values are computed at fixed dJ/d$\Omega$ values, as described above for s-wave $\bf{left}$, p-wave $\bf{middle}$, and d-wave  $\bf{right}$. Figure 4 shows the same, except for major axis fits to contours set at 20 percent of the peak emission of each halo.  

In the cases of s,p and d-wave, we find a significantly larger elliptical function in terms of its major axes and minor axes. This is expected, since in our previous results we showed in Chapter 2 that full baryonic physics enhances and increases the signal particularly its spatial extents. In many cases, the shape of the ellipse is rounder for the FIRE emission signals than it is for the DMO. 



Figures 5 and 6 plot the axis ratios (minor/major) obtained for the contour fits in DMO (vertial) versus FIRE (horizontal) for s-wave, p-wave, and d-wave, respectively. Figure 5 is produced from contour values in Table 1, while figure 6 is taken from peak 20 percent values in Table 2. The solid blue lines show the one-to-one relation to guide the eye. When points lie below the line, it means the FIRE runs are rounder on the sky. We see that, typically, the FIRE runs are indeed rounder. Though not always.  Importantly, the J-contours in FIRE have axis ratios that are similar, with a full range spanning $0.65-0.9$, and typical values around $\sim 0.8$.  The DMO runs show much larger variance $\sim 0.4 - 0.95$.

\section{Axes Ratios at Specific Contour Values}

We have consistently found that FIRE halos have a much greater extent, particularly for p and d waves. Here we also show the ratios of major axes for FIRE/DMO halos. As can be seen in figures 5 and 6, The ratios of semi- minor to semi- major axes show a greater range than the same ratios when the contour is chosen at the 20 percent of peak value for each halo. Here we see that even for the case of fixed levels for s,p, and d-wave DJ/DO, the majority of FIRE halos approach approx. 0.8 for s wave, with notable outliers m12m. Interestingly, m12m comes much closer to approx. 0.8 for p-wave and d-wave.

\section{Conclusions}

Using high-resolution zoom cosmological simulations of Milky-Way analogs described in \cite{2022MNRAS.513...55M}, we have constructed sky maps  of  dJ/D$\Omega$ for s,p and d-wave dark matter annihilation models and explored their shapes on the sky. Examples of these maps are shown in Figures 1 and 2. In order to quantify the shape of the emission on the sky, we have fit ellipses to contours of constant dJ/D$\Omega$ emission. From this analysis we find several important results. First, we observer fully self-consistent galaxy formation FIRE runs produce fairly circular emission contours, with the ratio of semi-minor to semi-major axes is typically $\sim 0.8$ for all three annihilation models considered. The full range of ratios is $0.65-0.87$. 

These results can be directly compared to Galactic Center Excess signals from Gamma Ray telescopes to determine if any of these annihilation models are consistent with the Galactic center excess signal or not. We find that  less complete, DMO runs display a much larger range of shapes on the sky ($0.4 - 0.95$) and are typically more elliptical than their FIRE counterparts. This result extends work done by \cite{grand2022dark} , who found that simulations containing gas and stars as well as dark matter are more morphologically consistent than DMO runs. Here we report new results for p and d wave as well as s-wave.

\cite{piccirillo2022velocity} have examined the velocity dependent J-factors for p and d wave, as well as Sommerfeld models using the Auriga simulations. Our results are fairly consistent with their findings. However,we have been able to provide more detailed information about the specific shapes of the emission signals on the sky which can be useful in determining the validity of s,p,and d wave models when comparing data from FermiLAT and other Gamma Ray Telescopes. This is due to the higher resolution of FIRE simulations.

For FIRE runs, we find that the major axis of the J-factor contours align closely with the galactic plane, within a few degrees for all cases. This can also be compared to a directly detected gamma ray signal from galactic centers to see if there is a consistent match or not.These results provide new expectations for the shape of dark matter annihilation emission signals. Specifically, if analysis priors were based on DMO expectations (which is standard) we would assume semi-minor to semi-major axes with a range of values for s,p,and d-wave emission (see Figures 3,4,5 and 6.) 
On the contrary, for the more accurate FIRE halos, the same ratios for the axes have a much narrow range of expected values near $\sim 0.8$. This is critical, as it shows a new expectation for what the expected signal would be in terms of its shape. 

We also find that the major axis of the J-factor maps for FIRE halos are always aligned with the galactic plane within a few degrees, meaning that excess emission out of the plane would be hard to explain with a dark matter annihilation signal.  We have provided the angles measured with respect to the galactic plane in tables 1 and 2.

There has been some debate in the literature on the shape of emission in the Galactic Center Excess. Ref. \cite{Macias18} find that the excess is described by a ``boxy bulge" shape, with short to long axis ratio (in our language) $\sim 0.55$.  This is quite flattened compared to our best-derived expectations from FIRE simulations (though not outside the realm that would have been expected from our DMO simulations, for example \texttt{ThelmaDMO} is more flattened than this.).  Indeed Ref.\cite{Abazajian20} have used this result to rule out a large class of thermal s-wave WIMP models.  Interestingly, however,\cite{di2021characteristics} found that the Galactic Center Excess has a shape on the sky well fit by an ellipsoid with a fairly round axis ratio $\sim 0.8 - 1$.  This shape of excess would be more easily explained by the shapes we report here.

Our theoretical results are in broad agreement with previous work, though no one has previously presented the same analysis and had comparable resolution. \cite{bernal2016spherical} used a J-factor-weighted inertia tensor over the whole sky and found that fully hydrodynamic halos tended to produce systematically more circular shapes on the sky than dark matter only versions.  Though not defined in precisely the same way, they report typical values of $\sim 0.8$ for the associated axis ratios, in good agreement with our work.
Future indirect searches for annihilating dark matter in the Milky Way, M31, and other galaxies will continue to rely on templates to model possible sources.  We have provided results that should be useful for dark matter templates here.

\section*{Acknowledgements}

We would like to acknowledge Professor James Bullock for his insights into shape fitting for the J factor Skymaps.

\section*{Data Availability}

The data supporting the plots within this article are available on reasonable request to the corresponding author. A public version of the GIZMO code is available at \href{http://www.tapir.caltech.edu/~phopkins/Site/GIZMO.html}{http://www.tapir.caltech.edu/~phopkins/Site/GIZMO.html}.



\bibliographystyle{mnras}
\bibliography{Velocity-Dependent-J-Factor.bib}





\bsp	
\label{lastpage}
\end{document}